\documentclass[reprint, prx, aps, notitlepage, nofootinbib,longbibliography,floatfix, superscriptaddress]{revtex4-1}
\usepackage{amsmath,amstext,amssymb,graphicx,bm,dcolumn,color,times, braket,caption}
\usepackage[colorlinks=true,citecolor=blue,urlcolor=red]{hyperref}
\usepackage[table]{xcolor}
\usepackage{graphicx}
\usepackage{environ}
\usepackage{amssymb}
\usepackage{amsmath}
\usepackage{amsthm}
\usepackage{amsfonts}
\usepackage{comment}
\usepackage{braket}
\usepackage{lineno}
\usepackage{subfig}
\usepackage{float}
\usepackage[normalem]{ulem}

\NewEnviron{resizeenv}{\resizebox{\linewidth}{!}{\BODY}}

\begin{document}

\title{Bell's shadows from satellites}
\author{Stav Haldar}\email{Current affiliation: Manning College of Information and Computer Sciences, University of Massachusetts Amherst, Amherst MA 01003. \\Email: stavhaldar@gmail.com}\affiliation{Department of Physics and Astronomy, Louisiana State University, Baton Rouge, LA 70803, USA}
\author{Rachel L. McDonald}\email{rmcdo33@lsu.edu}\affiliation{Department of Physics and Astronomy, Louisiana State University, Baton Rouge, LA 70803, USA}
\author{Sage Ducoing}\affiliation{Department of Physics and Astronomy, Louisiana State University, Baton Rouge, LA 70803, USA}
\author{Ivan Agullo}\affiliation{Department of Physics and Astronomy, Louisiana State University, Baton Rouge, LA 70803, USA}

\begin{abstract}
{Establishing reliable quantum links between a network of satellites and ground stations is a crucial step towards realizing a wide range of satellite-based quantum protocols, including global quantum networks, distributed sensing, quantum key distribution, and quantum clock synchronization. In this article, we envision a network of satellites and ground stations where quantum links are created through the exchange of entangled photon pairs.  We simulate the dynamics of a satellite constellation and a set of Bell tests between the constellation and ground stations. We identify the regions on Earth where Bell tests can be successfully conducted with a satellite or a set of them, at a specified level of confidence. These regions move with the constellation and will be referred to as {\em Bell violation shadows}. We demonstrate that these shadows provide valuable insights for the study and evaluation of many satellite-mediated or satellite-assisted quantum protocols.}
\end{abstract}
\maketitle
\section{Introduction}
\label{sec:Intro}

Bell's inequalities provide a powerful and convenient framework for testing quantum correlations \cite{RevModPhys.81.865}. In cryptography, they enable a device-independent test of quantum security \cite{PhysRevLett.97.120405}. In quantum networks, Bell tests serve to certify the usefulness of connections
\cite{Bancal2021, Wang2023}. Similarly, Bell tests play a crucial role in distributed sensing and distributed computing protocols \cite{Storz2023, PhysRevLett.121.220404, PhysRevLett.121.180505}. For applications such as quantum clock synchronization, where classical information is shared via quantum resources, Bell tests provide an added layer of quantum security \cite{Lee2019, Jozsa:2010, Jon_QCS_2018, white_paper}.

On the other hand, satellite-based quantum protocols will be essential for the scalable implementation of many of the technologies and applications mentioned above \cite{Sidhu2015, LSU_satsim, paper1}. By taking advantage of the low-loss communication channels provided by free space,
satellite constellations equipped with quantum hardware will complement and reinforce terrestrial systems designed to deliver quantum services. This motivates the central question addressed in this work: \emph{How can we reliably quantify the serviceable area and the quality of service provided on Earth by satellites carrying entanglement sources?}

Considering that the success of a wide range of applications is tied to the confidence with which Bell tests can be violated, we use this metric to assess the performance of entanglement distribution from satellites. Specifically, in this work, we identify the regions on Earth where Bell tests can be successfully performed with a certain level of confidence when entangled states are distributed from a network of satellites. We refer to these regions as the \emph{Bell violation shadows} of satellites or, simply, Bell shadows.

In any practical scenario, only a finite number of entangled bits (ebits) can be shared between two parties, meaning that Bell tests are successful only with a finite level of confidence. This limitation inherently leads to a sub-optimal success rate for any protocol that relies on these quantum correlations. Finite statistics play a critical role \cite{Sidhu2022}, particularity in long-distance and high-loss scenarios, such as satellite-to-ground or satellite-to-satellite quantum channels \cite{PhysRevLett.120.140405}. Bell shadows, therefore, become a valuable tool for analyzing and optimizing satellite-mediated quantum protocols. We expect our analysis to be relevant not only for designing protocols and quantum hardware for satellite payloads but also for informing the design of satellite constellations and quantum networks---a complex optimization problem in its own right.

\begin{figure*}[ht]
\includegraphics[width=0.3\linewidth]{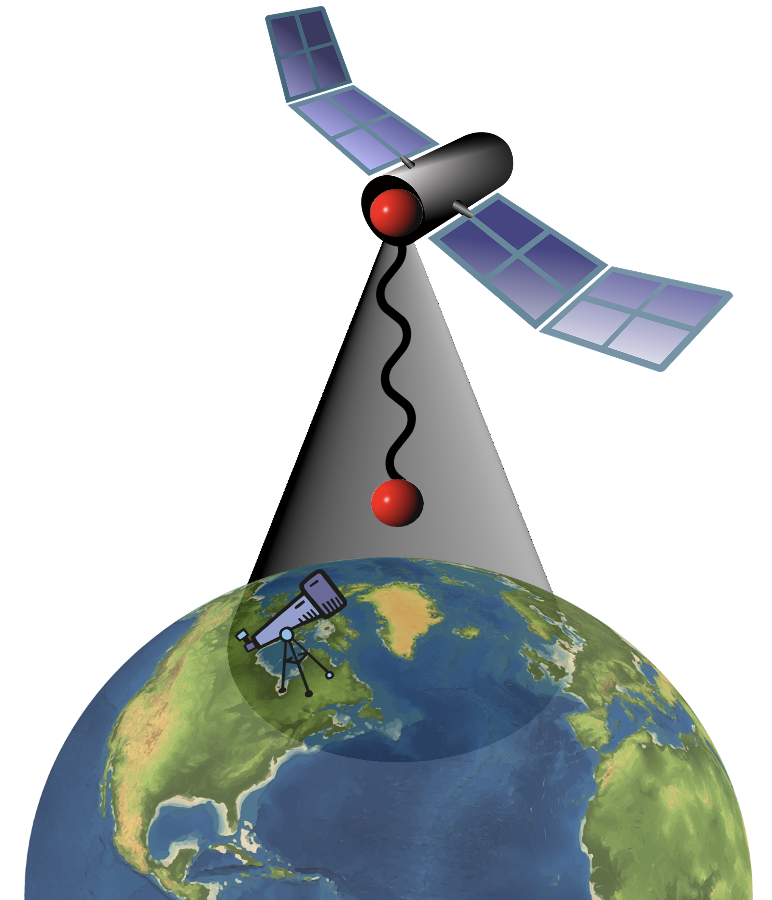}
\includegraphics[width=0.28\linewidth]{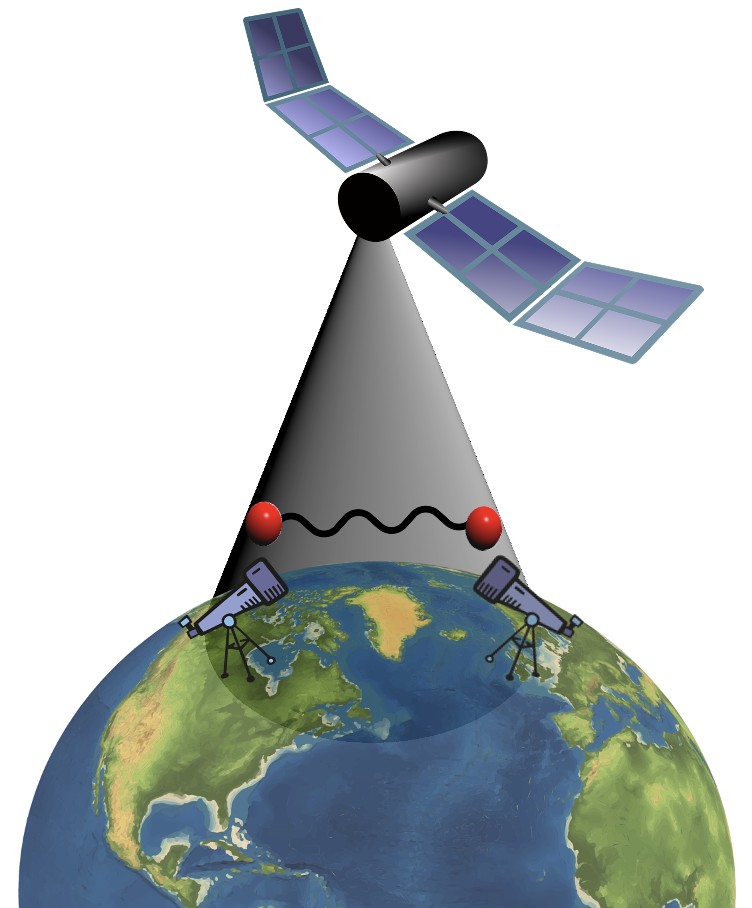}
\includegraphics[width=0.4\linewidth]{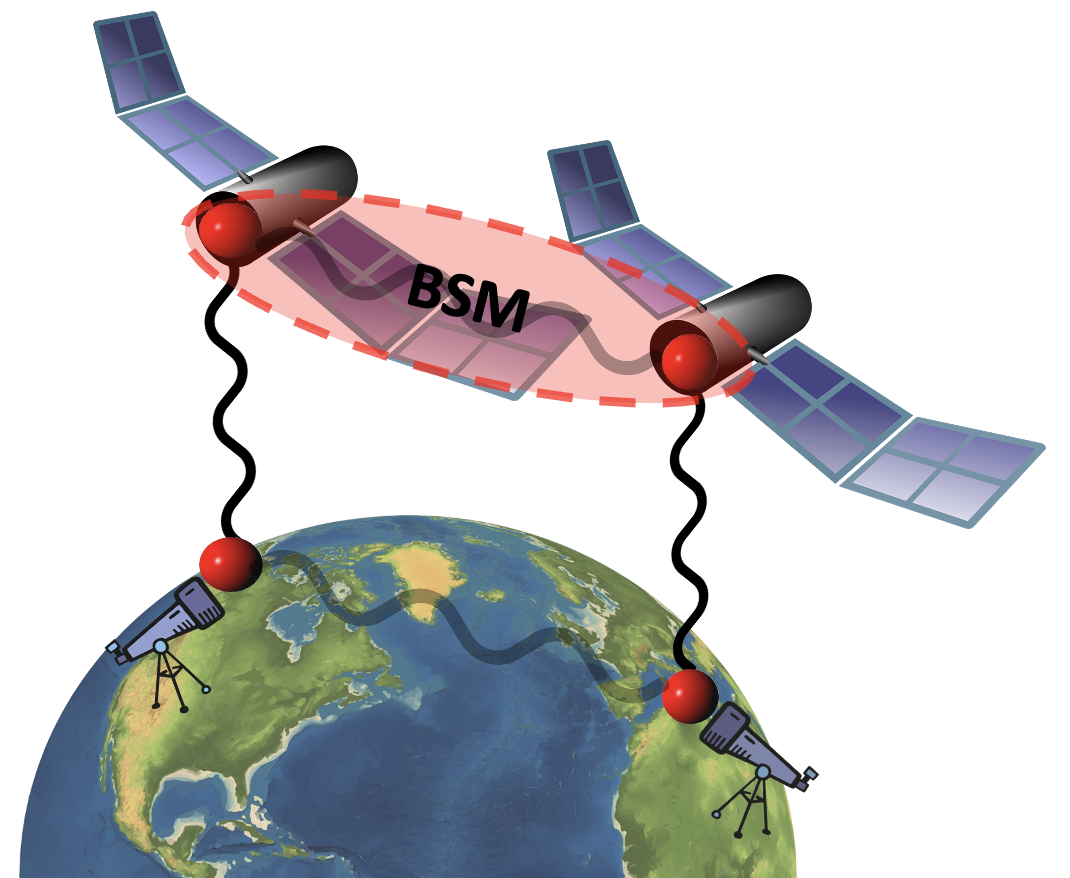}

\caption{Bell violation shadows for: (left) single downlink scenario where entangled photons are shared between a satellite and an optical ground station, with the satellite carrying the entangled photon source; (center) a double-downlink scenario where ebits are shared between two ground stations using an ebit source on the satellite; and (bottom) connected satellites with independent downlinks where satellites share ebits not only with ground stations but also among themselves. All these different scenarios are relevant for different quantum protocols such as key distribution, networks, and clock synchronization. We analyze the performance of such protocols under these scenarios using the concept of Bell violation shadows.
}
\label{fig:scenarios}
\end{figure*}

More specifically, in this paper we will quantitatively analyze  three satellite-based entanglement distribution scenarios: (1) Ebits are shared between a satellite and a ground station via downlinks (i.e, with the source located onboard a satellite). This scenario can be easily modified to study uplinks (source located at a ground station).
(2) Ebits are sent from a satellite to two ground stations, using an entanglement source onboard a satellite. A variation of this scenario, utilizing two independent downlinks of the type described in scenario (1), followed by an entanglement swapping operation at the satellite, is also discussed in section \ref{subsec:Quantum_Networks}.
(3) Connected satellites: 
 Satellites share entanglement among themselves and with ground stations. By acting as quantum repeaters, they can extend the scale of the network. These three scenarios are sketched in Fig. \ref{fig:scenarios}.

In existing literature, the quality of entanglement distribution via satellite constellations has primarily been evaluated using ebit rates \cite{LSU_satsim, paper1, rate_based1, rate_based2}. Quality of service is enforced by setting a cut-off rate or threshold below which the distribution task is deemed unsuccessful \cite{paper1}. Bell shadows offer improvements over the ebit-rate or cut-off-based approach in two significant ways.

On the one hand, the choice of ebit rate threshold is highly dependent on the protocol. While a higher cut-off corresponds to demanding higher quality of service, it does not necessarily guarantee the presence of quantum correlations. For instance, only in the case of a simple pure bosonic loss channel does the ebit rate directly correspond to the (Werner) state fidelity and, consequently, to Bell violations. In more realistic scenarios involving noise---such as background counts, dark counts, or atmospheric birefringence (some of which are considered in this work)---there is no direct correlation between nonzero ebit rates and the presence of genuine quantum correlations (see section ~\ref{sec:sim_results}). Conversely, a violation of a Bell test guarantees the presence of quantum correlations, even in noisy conditions \cite{Horodecki1995}.
On the other hand, in realistic settings where the time available for data acquisition is finite, both ebit rates and state fidelity are subject to fluctuations. Ensuring reliable quantum correlations under such conditions requires consideration of the variance of Bell violations (the CHSH number). Unlike arbitrary cut-off rates, Bell violation shadows define regions of Earth where Bell violations occur with statistical significance, accounting for both the effects of loss and noise (e.g., drops in raw ebit rates and state fidelity) and the effects of finite statistics (fluctuations in rates and fidelity).
It is worth noting that Ref.~\cite{Sidhu2022} specifically examines the impact of finite statistics on quantum key rates derived from satellite-based entanglement sources. However, our results are aimed at providing a more versatile figure of merit that also assesses network scalability, distinguishing them from those in Ref.~\cite{Sidhu2022}. Furthermore, Ref.~\cite{LEO_exact_downlink} also considers  violations of Bell inequalities in the context of a LEO-satellite single-downlink scenario, focusing primarily on quantum key distribution.

The structure of the paper is as follows: In Section~\ref{sec:Bell_shadow}, we introduce and discuss Bell shadows for satellites in the  scenarios discussed above. In Section~\ref{sec:simulation}, we briefly describe our simulation techniques and the simulation code we use to dynamically evaluate these shadows in the presence of loss, background noise, and dark counts. Our simulation results are presented in Section~\ref{sec:sim_results}. More specifically, we look at the different factors on which the extent of Bell violation shadows depends, which include confidence level, acquisition time, background, and dark count rates. We look at several scenarios relevant for practical applications, such as asymmetric loss and noise. In Section~\ref{sec:apps}, we provide further simulation results as examples of the utility of Bell violation shadows in different applications. These applications, including  quantum key distribution, quantum clock synchronization, and quantum networks, exemplify how Bell shadows can be used to assess the performance of a variety of protocols. Finally, in Section~\ref{sec:conclusion} we present our conclusions and discuss future work. 

\section{Bell violation shadows}
\label{sec:Bell_shadow}

Satellite-based entanglement distribution using free-space communication channels enables the sharing of quantum resources over significantly larger distances---network scale---compared to traditional fiber-optic channels. Given that this is the primary advantage of satellite-based protocols, it is essential to evaluate the network scales achievable through these methods.

Consider a satellite in a low Earth orbit having an entanglement source onboard,  aimed to share entangled photon pairs (ebits) with some ground station on Earth. One member of the pair is measured locally at the satellite while the other member is sent to the ground station, which will detect it with some probability. As mentioned above, we quantify the success rate of a quantum protocol involving a satellite and a ground station using the probability that a Bell test is successful. Specifically, consider the CHSH number, defined as 
\begin{eqnarray}
    S = E(a_1,b_1) + E(a_1,b_2) - E(a_2,b_1) + E(a_2,b_2)
    \label{eqn:CHSH}
\end{eqnarray}
where $E(a_i, b_j)$ is the expectation value of the product of measurement outcomes (in our case, a polarization measurement, with possible outcomes 
$\pm 1$), when measurements by Alice and Bob (a and b) are made in the $a_i, b_j$ bases, respectively ($i,j \in \{1,2\}$). For the CHSH number to achieve its maximum (absolute) value, the bases sets $a = \{a_1, a_2\}, b = \{b_1, b_2\}$ must be chosen optimally. 
For example, for the maximally entangled Bell states of two photons entangled in their polarization degrees of freedom, the optimal choice is as follows: $a = \{\Pi_{0^\circ}, \Pi_{45^\circ}\}$ and $b = \{\Pi_{22.5^\circ}, \Pi_{-22.5^\circ}\}$, where $z$ is the direction of propagation and $\Pi_\theta$ denotes the projector on the linearly polarized state making angle $\theta^\circ$ with the $x$-axis in the $x$-$y$ plane. For this basis set $S$ achieves its maximum possible value of $2\sqrt{2}$.

Furthermore, for each individual measurement, Alice will randomly choose one of the two operators in the set $a = \{\Pi_{0^\circ}, \Pi_{45^\circ}\}$; similarly, Bob will independently and randomly choose one of the operators in the set $b = \{\Pi_{22.5^\circ}, \Pi_{-22.5^\circ}\}$. In a realistic situation with finite statistics, the CHSH number will deviate from around the theoretically computed value \cite{Elkouss2016}. 

In this situation, the question of a successful Bell test must be defined more carefully. This can be done as follows. Consider that photons are collected during a finite acquisition time $t_{acq}$, and their polarization is measured by both parties in the way indicated above. The CHSH number $S$ is calculated using these measurements. We call this one round or one run of the Bell test. We will motivate the choice of $t_{acq}$ later, but for now it suffices to consider this to be an experimental parameter. Since pair-production from an entanglement source (like SPDC) is a probabilistic process, the exact number of photons received in any run is random (it follows a Poisson distribution in our simulations).

Our goal is, given a fixed value of $t_{acq}$, to characterize the probability distribution of $S$, particularly its mean value $\bar S$ and standard deviation $\sigma_S$. With this information, we can determine the locations on the Earth's surface where $S$ is expected to be larger than 2 with some confidence level. More specifically, we will say that, given a satellite and a certain location on Earth, a CHSH test will be successful at ``$n$-sigmas'', when 
\begin{equation} \bar S-n\, \sigma_S\geq 2. \end{equation}

Because losses scale with the distance between the satellite and the ground station, the region right beneath the satellite will correspond to the largest confidence level. Furthermore, such a region is ``anchored'' to the satellite, following it and moving along the Earth's surface as the satellite moves. These are the Bell shadows discussed above. 

Thus, the computation of Bell shadow reduces to the calculation of the probability distribution $P(S)$ of the CHSH number $S$. This calculation involves several difficulties. 

On the one hand, this probability distribution depends on a large number of parameters, including $t_{acq}$, the satellite's orbit, the emission rate of the source of entangled photons on board the satellite, background noise, dark count rate, atmospheric losses, etc. Furthermore, these parameters vary considerably over time, adding significant difficulties to the evaluation of the probability distribution---particularly, when considering a large number of satellites and ground stations, as we do below. Additionally, the evaluation of the probability distribution requires knowledge of the quantum-mechanical state after it has passed through a lossy and noisy channel. The effects of loss and noise are probabilistic in nature. The state of the photon pair is therefore described by a time-dependent statistical mixture, such as a Werner state or other forms of average density matrices, although such a description is accurate only in the asymptotic case of infinite measurements, which is precisely the assumption we wish to avoid.  

In Appendix \ref{app:A}, we provide additional details about the analytical computation of $P(S)$ and the associated practical challenges. These challenges are particularly severe in the complex satellite networks analyzed in this article.  

Given these difficulties, we adopt a different strategy. Instead of pursuing a purely analytical computation of $P(S)$, we use computer simulations to model the dynamics of satellites and the Bell tests. These simulations inherently account for all the effects mentioned earlier, including the time-dependent variations induced by satellite trajectories and the finite statistical effects. This approach provides a direct and reliable method for obtaining the Bell shadows under investigation.  

We also note that this simulation-based strategy based on confidence levels, can be further  formalized using hypothesis testing frameworks, as detailed in Refs.~\cite{Elkouss2016, Bierhorst2015}.

Specifically, we proceed as follows. We start by making a choice of $t_{acq}$, the time span of each run. We choose $t_{acq}$ in the range $t_{acq} \approx 0.1-10 \,\rm{ms}$. This guarantees that enough e-bits are shared between the two parties, keeping in mind that  only a very small fraction of the  ebits shared will actually be used for verification of Bell tests. At the same time, continuous verification is also crucial in dynamic settings, such as satellite-based entanglement distribution. Thus, $t_{acq}$ should be small enough so that noise and loss levels do not vary too much within one run.

Choosing $t_{acq}$ within this range yields, for the values of pair-creation rate, losses, and dark counts used in this analysis (specified in the next section), practically useful rates of ebits received by both parties of approximately $100 \, \mathrm{to} \, 10000 \,\mathrm{Hz}$.  

Our computer simulation recreates the dynamics of the satellites, the random generation of entangled photon pairs (with a Poisson distribution) at the satellites, the propagation from  satellites to ground station through a lossy and noisy channel, and the detection procedures at both stations and  satellites. For each detected pair, the polarization bases are randomly chosen at both the satellite and the ground station as specified above, and the CHSH parameter $S$ is computed by averaging over all pairs detected during the duration $t_{acq}$ of the run.  

We then concatenate several runs sequentially and compute, for a pair satellite-ground station, the mean value $\bar{S}$ over all runs and the mean squared deviation $\sigma_S$. For all simulation results presented in this paper, the number of runs is fixed at $30$. During these runs, the distance between the satellite and the ground station varies. Although for $t_{acq}$ in the range $0.1-10 \,\mathrm{ms}$ and $30$ runs this variation is small, our simulations account for these changes.

 We conclude this section providing two visual examples of Bell shadows. These examples aim at  equipping the reader with useful intuition before describing details of our simulation methods in the next section. 
 
\begin{figure}[ht]
\includegraphics[width=0.95\linewidth]{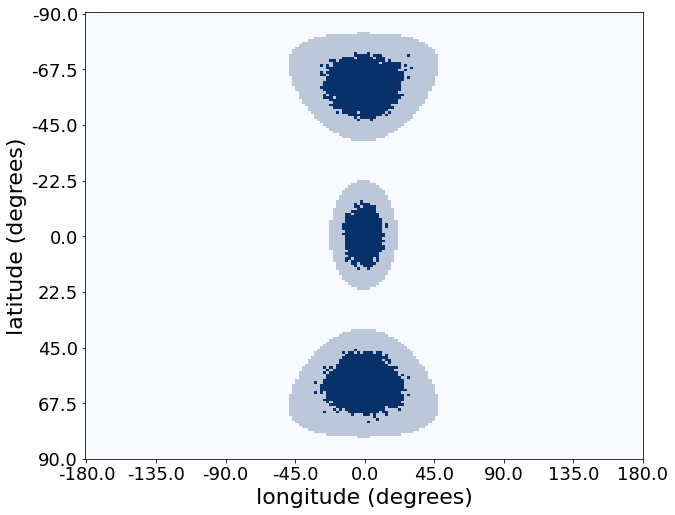}
\caption{Snapshots of Bell  shadows for a downlink scenario. The satellite is in a polar orbit at 500 km altitude. Each shadow patch is a snapshot when the satellite is just overhead some point on Earth on the Prime Meridian (the central shadow is when the point is  also at the Equator). Acquisition time is set to 1 ms and we consider a 1-$\sigma$ confidence level. The source rate is 10 million ebits/s and background (at the ground station) and dark count rates (for satellite detectors) are 10 kHz and 1 kHz respectively. The outer shadow is the visibility shadow of the satellite. Since Bell violation shadows are smaller than the visibility shadows, just being visible to the satellite does not guarantee the availability of verifiable quantum correlations in a rate-restricted scenario relevant to most quantum protocols.}
\label{fig:single_shadow}
\end{figure}

The first example consists of a single-downlink scenario, in which the source of entangled photons is on board the satellite.\footnote{ This scenario can be adapted to include uplinks, i.e., when the ground station has the ebit source. The Bell shadow in such a scenario still quantifies the region on Earth where, if a ground station is located, it can share quantum correlations reliably with a satellite. Uplink shadows are generally smaller than downlink shadows due to larger losses.} One member of the pair is locally measured at the satellite, while the other photon is sent to a ground station on the Earth. In this example, the satellite follows a polar orbit of 500 km altitude. Fig.~\ref{fig:single_shadow} shows three Bell shadows, computed at one $\sigma$, corresponding to three different instants in the satellite trajectory.  We also draw the visibility shadows in light blue---the region on the Earth's surface from which the satellite is visible at an instant---
in order to point out the difference between being able to see the satellite and receiving quantum correlated states with high confidence (dark blue region). As one moves farther from the center of the shadow, a smaller number of ebits are successfully shared between the ground station and the  satellit, thus leading to less confidence in the mean CHSH number and a low probability of a successful Bell test. The sparseness at  edge of the Bell shadow is an effect of finite statistics. 

The second example consists of a double-downlink scenario---entangled photons are generated on the satellite, and each member of the pair is sent to a ground station. 
This scenario requires a modified definition of Bell shadow. Two locations on the Earth are part of the shadow if they can perform a Bell test with a high enough success rate using ebits distributed from the satellite. In a double-downlink scenario, the shadow of a satellite must be determined by first fixing one of the ground stations. Thus, the Bell violation shadow is defined with respect to a satellite and some chosen location on Earth. 

\begin{figure}[ht]
\includegraphics[width=\linewidth]{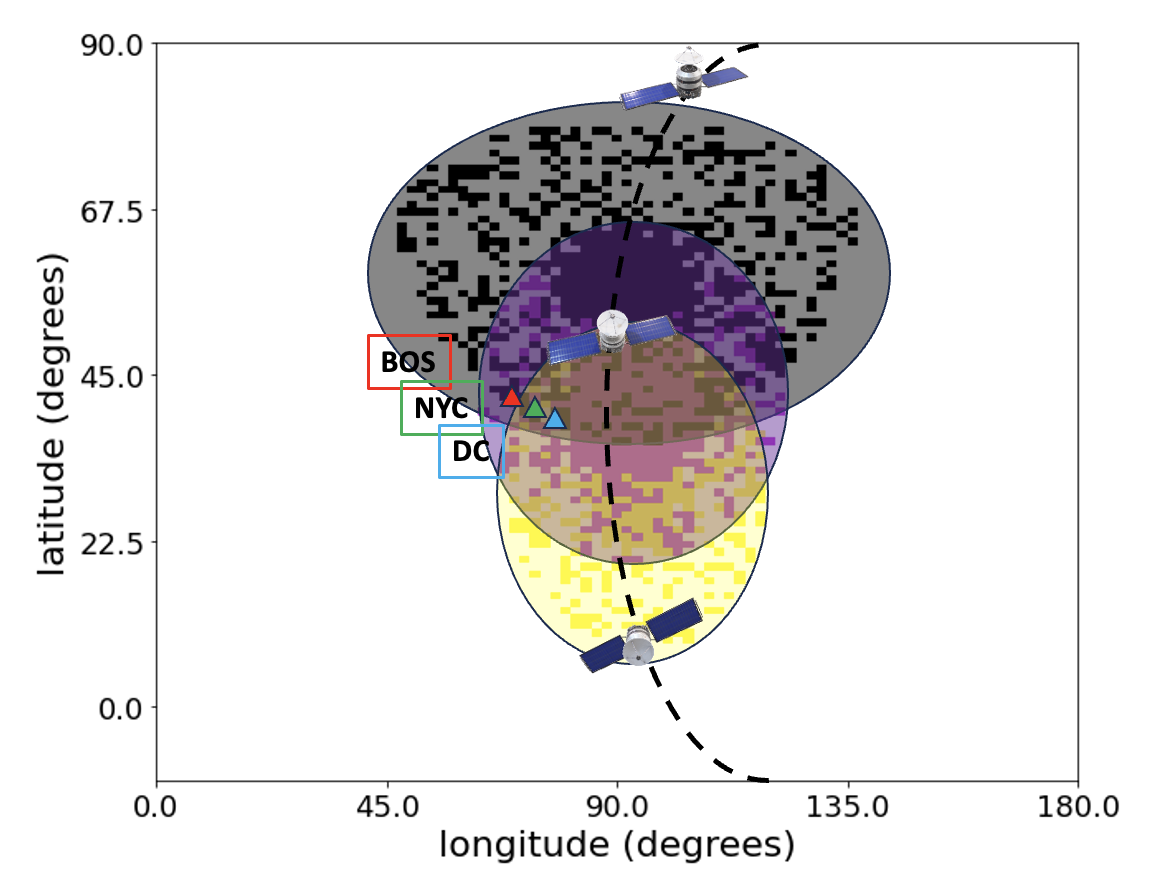}
\caption{Snapshots of Bell violation shadows for a double-downlink scenario. The satellite is in a polar orbit at 500 km altitude. Each shadow patch is a snapshot of the satellite flying by over close to New York City. Ellipses denote visibility shadows, while spots of different colors denote locations at which Bell tests are successful at one-$\sigma$ confidence. The total time elapsed in this shadow track is 450 seconds, which is close to the visibility period from NYC. In Sec.~\ref{subsec:QKD}, we consider a quantum key distribution protocol using this configuration and look at the performance of a network consisting of New York City, Boston and Washington D.C.}
\label{fig:dd_shadow}
\end{figure}

Fig.~\ref{fig:dd_shadow} shows the Bell violation shadows for such a double-downlink scenario with New York City (NYC) as the fixed point on Earth. As the satellite passes close to NYC, NYC and other cities near it pass through different sub-regions within the shadow. We will later show the role this plays in determining the performance of a concrete quantum protocol, namely quantum key distribution (QKD) (see Sec.~\ref{subsec:QKD}). 

Note that the Bell shadow in the double-downlink scenario is generically not symmetric with respect to the fixed location. It is symmetric only when the fixed ground station is directly underneath the satellite. This is a consequence of the fact that losses depend on the distance traveled by the photons. In fact, in some circumstances, the fixed ground station may even be outside the Bell shadow---it only needs to be visible from the satellite. This happens when the fixed ground station is close to the edges of the visibility shadow; in that case, the reference ground station will be unable to perform a successful Bell test with nearby ground stations. However, it will be able to perform a successful Bell test with ground stations closer to the point  right underneath the satellite, which suffer from fewer losses.

In Section \ref{subsec:Quantum_Networks}, we will  consider a third scenario  (see Fig.~\ref{fig:q_net_compare_constellation}) where satellites in a constellation can share entanglement among themselves. In this case, the union of the Bell shadows of individual satellites can be considered a single Bell shadow associated with the entire constellation. Acting as repeaters, the satellites can allow two cities which are not  visible simultaneously from any individual satellite in the constellation to share ebits, via entanglement swapping operations on board satellites.

\section{Simulation methods}
\label{sec:simulation}
For our simulations, we consider entangled photon pair sources (SPDC sources) with fixed rates and utilize free space communication channels to distribute them. The hardware parameters which are kept fixed for all our simulations are shown in Table~\ref{tab:table1}.
\begin{table}[h]
\centering
\begin{tabular}{|l|l|}
\hline
Altitude of satellite                              & h = 500 km         \\
\hline
Operational wavelength                            & $\lambda$ = 810 nm         \\
\hline
Radii of telescopes                  & ($r_{sat}$, $r_{gs}$) = (10 cm, 60 cm) \\
\hline
Detector efficiencies                        & ($\kappa_{sat}$,$\kappa_{gs}$) = (0.5, 0.5)     \\
\hline
Source rate                                       & $R$ = $10^7$ entangled-pairs/s   \\
\hline
Background noise rate                                       & $R_{\rm{bkg}}\in [1,100]$ kHz   \\
\hline
Dark count rate                                     & $R_{\rm{dc}} \in [0.1,10]$ kHz \\
\hline
\end{tabular}
\caption{Hardware parameters used for the simulations. $r_{sat}$, $r_{gs}$ are satellite and ground station telescope radii, respectively.}
\centering
\label{tab:table1}
\end{table}

The simulation consists mainly of three parts: (1) simulating the dynamics of the satellite and ground station, (2) simulating the lossy quantum communication channel between them, and (3) randomly generating and measuring the polarization of the photons at the relevant parties (ground station or satellite), including background noise and dark counts.
At every time step, we update the positions of the ground station and the satellite to evaluate the link distance for the photons that are produced in that time step. The link distance enters as an input into the efficiency $\eta$ of the free space communication channel. Furthermore, the visibility of one party with respect to another is also calculated from their position vectors. For a ground station-satellite pair, the angle between the ground station zenith line and the line joining it to the satellite must be less than $\pi/2$ to ensure visibility.

We take into account the various losses by calculating effective efficiency factors. We must distinguish between the uplink (ground station to satellite) and downlink (satellite to ground station) $\eta_{(up)}$ and $\eta_{(down)}$ respectively, since we assume different telescope radii for the two parties. Atmospheric turbulence and scattering effects would also be slightly asymmetric, but our model does not include these details. 

$\eta_{(down)}$ is the probability that a photon generated out of the SPDC source at the satellite will lead to a double detection event, i.e., one partner will be detected at the satellite locally and the other at the ground station after traveling through free space and the atmosphere. Similarly for $\eta_{(up)}$.
We model the satellite-ground station quantum communication channel as follows. For concreteness, let us focus on a downlink channel. The following will hold similarly for an uplink channel. We assume clear skies and approximate the downlink channel as only lossy (background noise is accounted for at the detectors). That is, photons are either transmitted through the channel or lost in transmission. The dominant sources of loss are (1) beam spreading (free-space diffraction loss), (2) atmospheric absorption/scattering, and (3) non-ideal photodetectors on the satellite and on the ground. We characterize these loss mechanisms by their transmittance values, which is the fraction of the received optical power to the transmitted power (which is also equal to the probability of transmitting/detecting a single photon). Let these transmittance values be, respectively,
\begin{equation*}
  \eta^{(down)}_{fs}(L), \quad \eta^{(down)}_{atm}(L), \quad \kappa_{sat}, \quad \kappa_{gs}\, ,
\end{equation*}
where the superscripts refer to the downlink, $fs$ refers to free-space diffraction loss, $atm$ to atmospheric loss, $L$ is the link distance (physical distance) between satellite and receiver (which in turn depends on the satellite altitude, $h$, position of the satellite in its orbit, and position of the ground station on Earth's surface), $\kappa$ denotes non-ideal detection efficiencies for the onboard satellite detector array ($sat$) as well as the detector array at the ground station ($gs$), and all transmittance values are less than or equal to $1$. Simple analytic formulae are used to estimate the free-space and atmospheric transmittance values in accordance with \cite{LSU_satsim}.
Therefore, the overall efficiencies are given as:
\begin{eqnarray}
\label{eqn:39-40}
    \eta_{(down)} &=& \eta^{(down)}_{fs}\eta^{(down)}_{atm}\kappa_{sat}\kappa_{gs}\, , \\
    \eta_{(up)} &=& \eta^{(up)}_{fs}\eta^{(up)}_{atm}\kappa_{sat}\kappa_{gs}\, . 
\end{eqnarray}
Every photon pair that is generated from the SPDC source is assigned a polarization (horizontal or vertical). A standard Monte Carlo procedure is used to generate all probabilistic events. Lets say a photon pair is generated. We consider it to be the ideal singlet state $\ket{\psi^-} = \frac{1}{\sqrt{2}}(\ket{HV} - \ket{VH})$. Now the two parties A and B are free to choose their basis (from a pre-decided set). If a double detection event occurs, the measurement outcomes are correlated with outcome probabilities dictated by the Born rule. If one partner is lost, of course, no contribution to Bell test occurs. Further, if one partner is lost but instead a background noise photon (generated with an independent constant rate at each receiving telescope) is generated at the receiving end, a randomly correlated outcome is produced, and these contribute to classical correlations. Similar statistics are generated when both ebit partners are lost but two background photons are generated. In cases when an ebit partner and a noise photon arrive at a receiver in the same time step, we treat the detection as a background photon or as a ebit probabilistically, with  the probability determined by the ratio of the background rate to the rate at which ebits are received. Dark counts at each detector, i.e. local and receiver, are included in a similar way as noise photons since they also lead to random/classical correlations. 

We re-emphasize that, from the perspective of extracting genuine quantum correlations, the use of Bell shadows is advantageous compared to demanding a minimum ebit rate exchanged between the two parties. This is because Bell shadows account for the deleterious effects of background noise and dark counts on entanglement, as well as all time-dependent effects arising from relative motion.

Finally, the Bell test statistics are generated. It is computationally efficient to use parallel processing to generate multiple independent runs of the Bell test and get the mean and standard deviation of CHSH number $S$. The number of runs and length of each run --- acquisition time --- is determined by practical constraints, such as rate requirement, the total visibility time of a satellite from a point on Earth, the fraction of the collected data that can be utilized for Bell tests, etc. In the next section, we study the effect of acquisition time, background noise and dark counts on the Bell violation shadows (fixing the number of runs to 30).

\section{Simulation results}
\label{sec:sim_results}
\begin{figure}[ht]
\includegraphics[width=\linewidth]{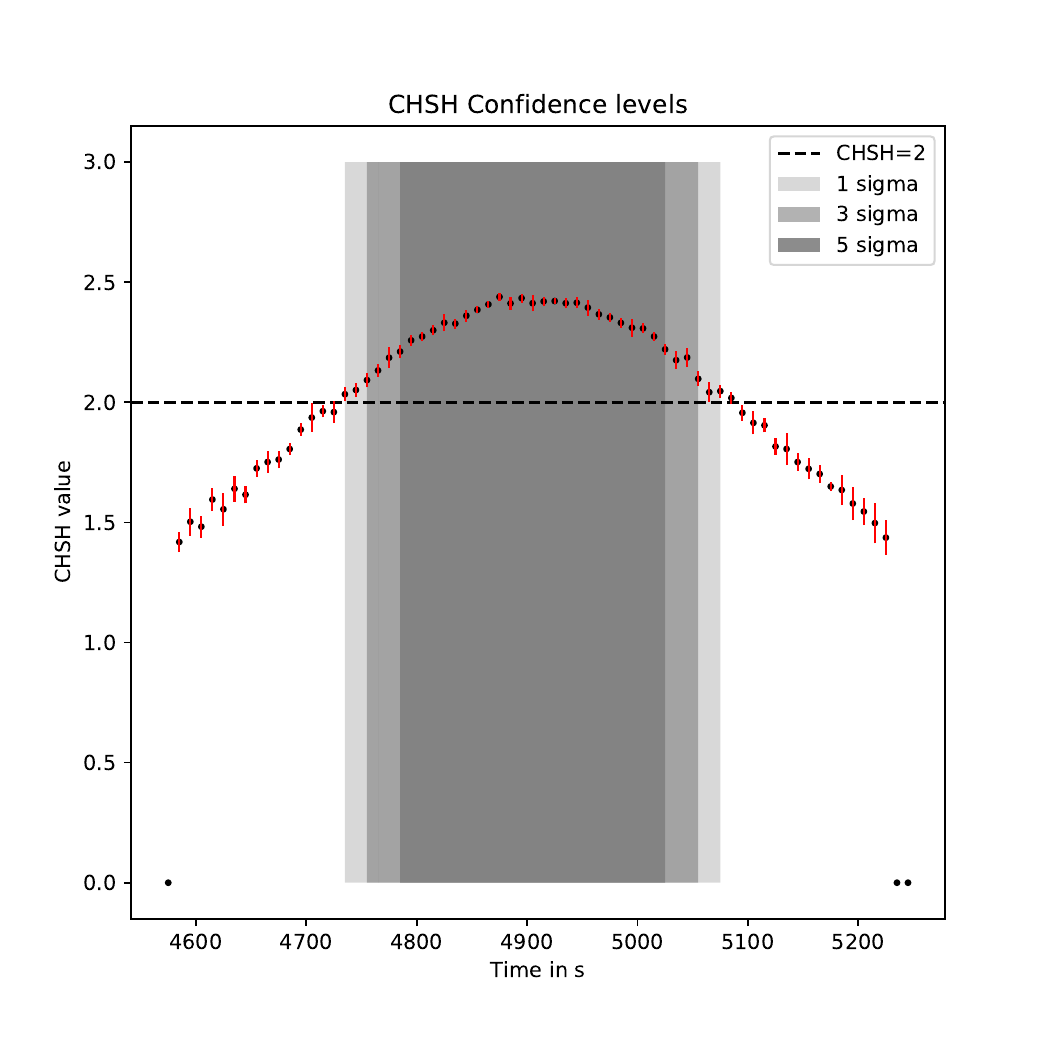}
\caption{
The average CHSH value during the visibility period of a satellite at 500km altitude following a polar orbit which passes through New York's zenith at $t\approx 4900$s, with a ground station located at New York City. Each black point corresponds to the value of the CHSH number at a given instant during the satellite's motion, and the red line on top of the points indicate 1-$\sigma$ deviations. The dark count and background rates used ar 100 Hz and $r_{\rm bkg} = 25$ kHz, respectively. $t_{\rm acq} = 1$ ms is used in this simulation. The time intervals during which the Bell test is successful with confidence levels 1-$\sigma$, 3-$\sigma$, and 5-$\sigma$ are shown with light, medium, and dark shading, respectively.}
\label{fig:CHSHstds}
\end{figure}

\begin{figure}[ht]
\includegraphics[width=\linewidth]{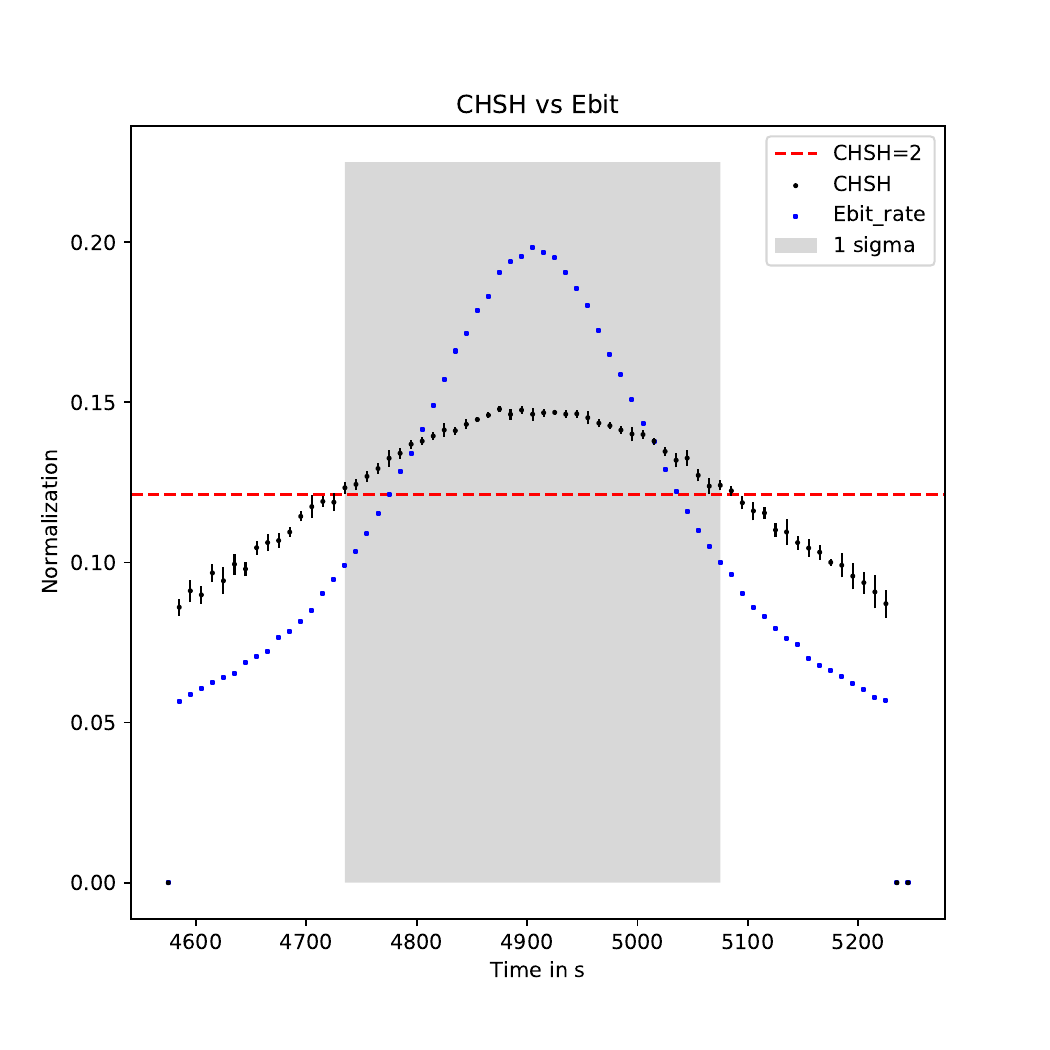}
\caption{The average CHSH value and ebit rate are displayed together, for the same parameters as in Fig.~\ref{fig:CHSHstds}. Each data set is normalized for clearer comparison. The shaded region shows when the Bell test is successful at the 1-$\sigma$ confidence level.}
\label{fig:CHSHvEbit}
\end{figure}

In this section, we present the results of our simulations. We consider first a single LEO satellite in polar orbit at 500 km altitude as a representative example. The results for the first scenario, i.e., the single downlink case, are presented first. To develop some intuition, in Fig.~\ref{fig:CHSHstds} we first show the CHSH number and its variance as a function of time. 
The observed time variation of the CHSH number has a simple explanation in terms of distances. Directly underneath the satellite, losses are minimal, hence ebits shared reach a maximum number. This translates to a high Bell violation and also a higher level of confidence on the test results. This can also be seen in Fig.~\ref{fig:CHSHvEbit}. 
At the same time, it is clear from our simulations that just having some ebit exchange rate is not enough for Bell's inequality to be violated, as can be seen near the tails of the curve in Fig.~\ref{fig:CHSHvEbit}.

\begin{figure*}[ht]
\centering
    Acquisition Time
\\
\includegraphics[width=0.32\linewidth]{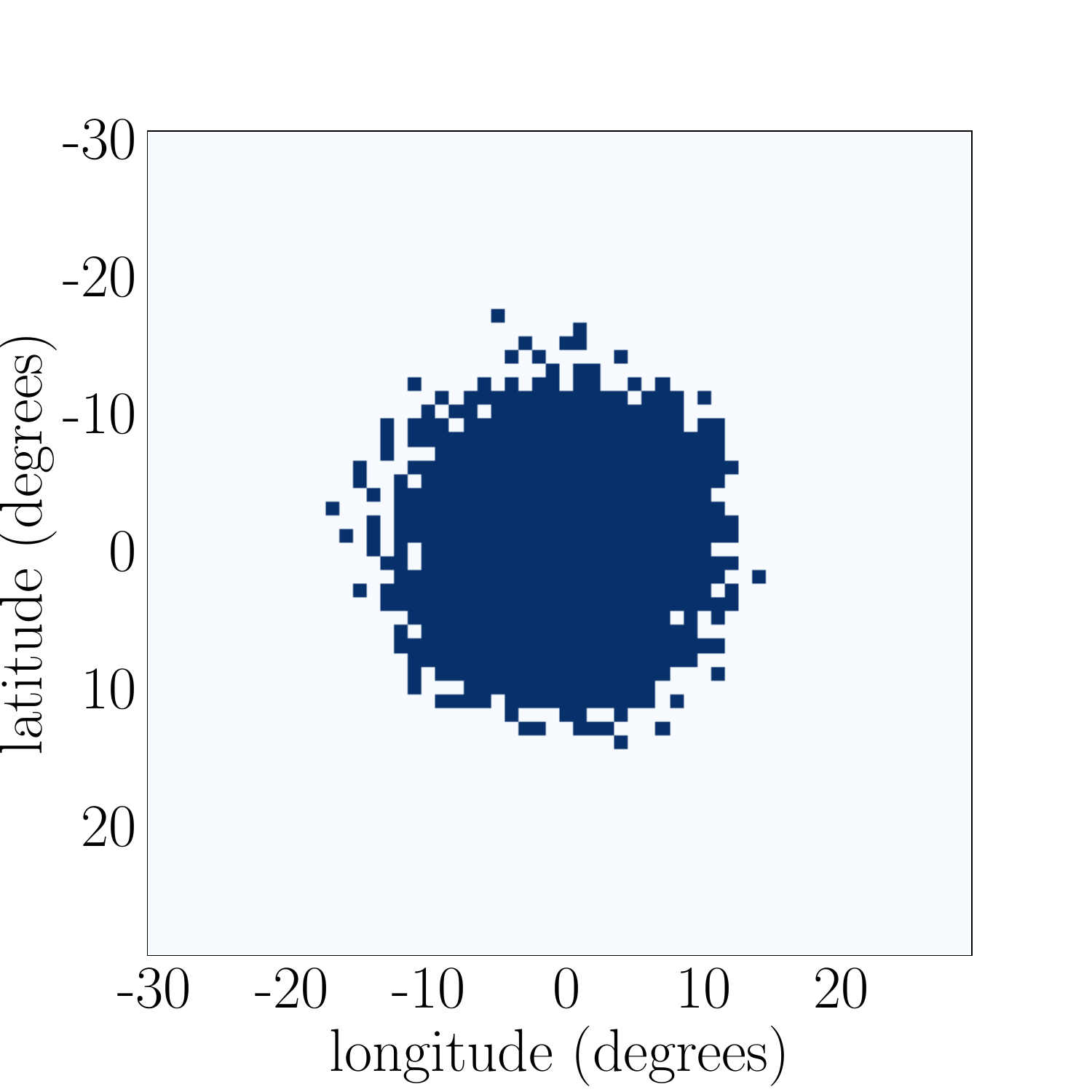}
\includegraphics[width=0.32\linewidth]{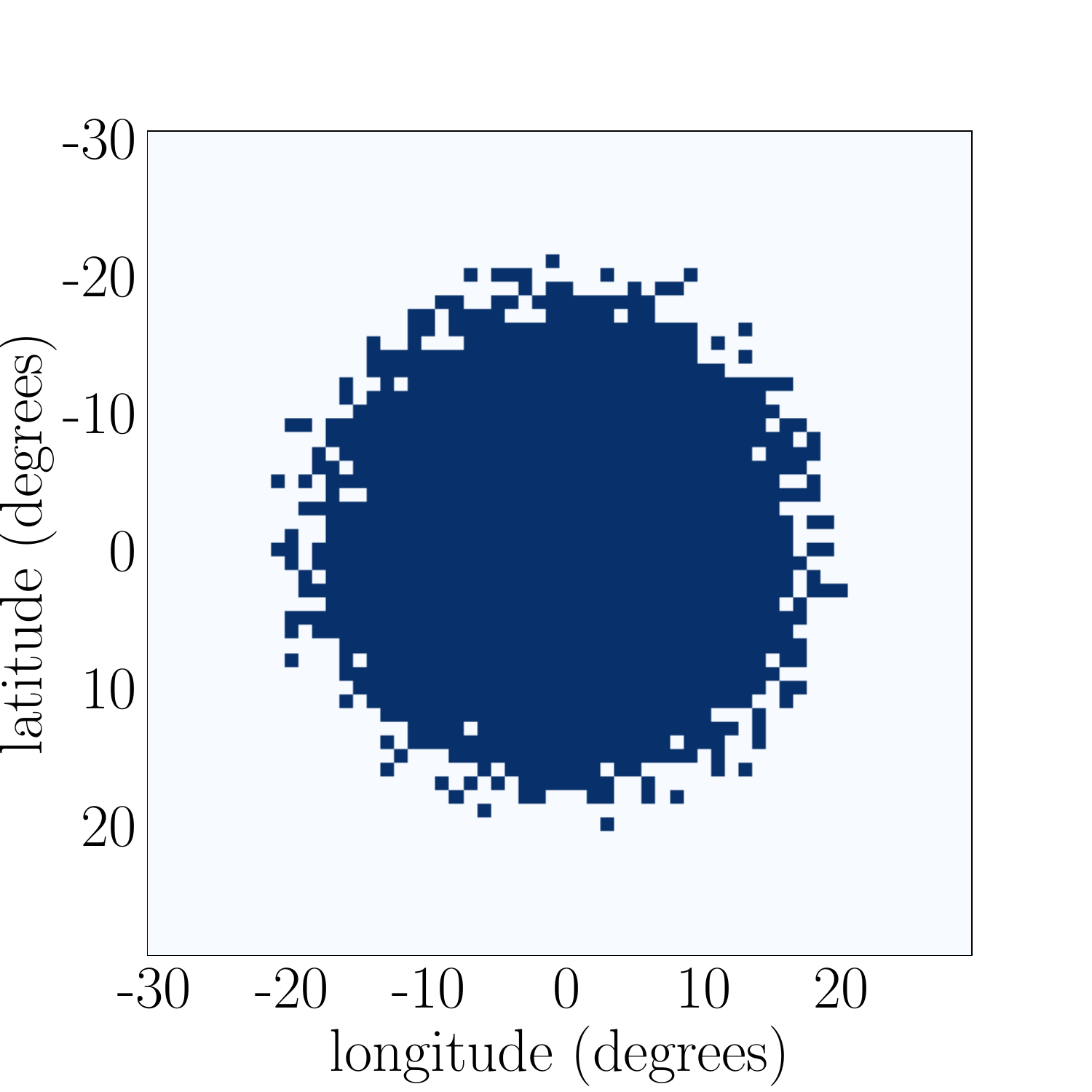}
\includegraphics[width=0.32\linewidth]{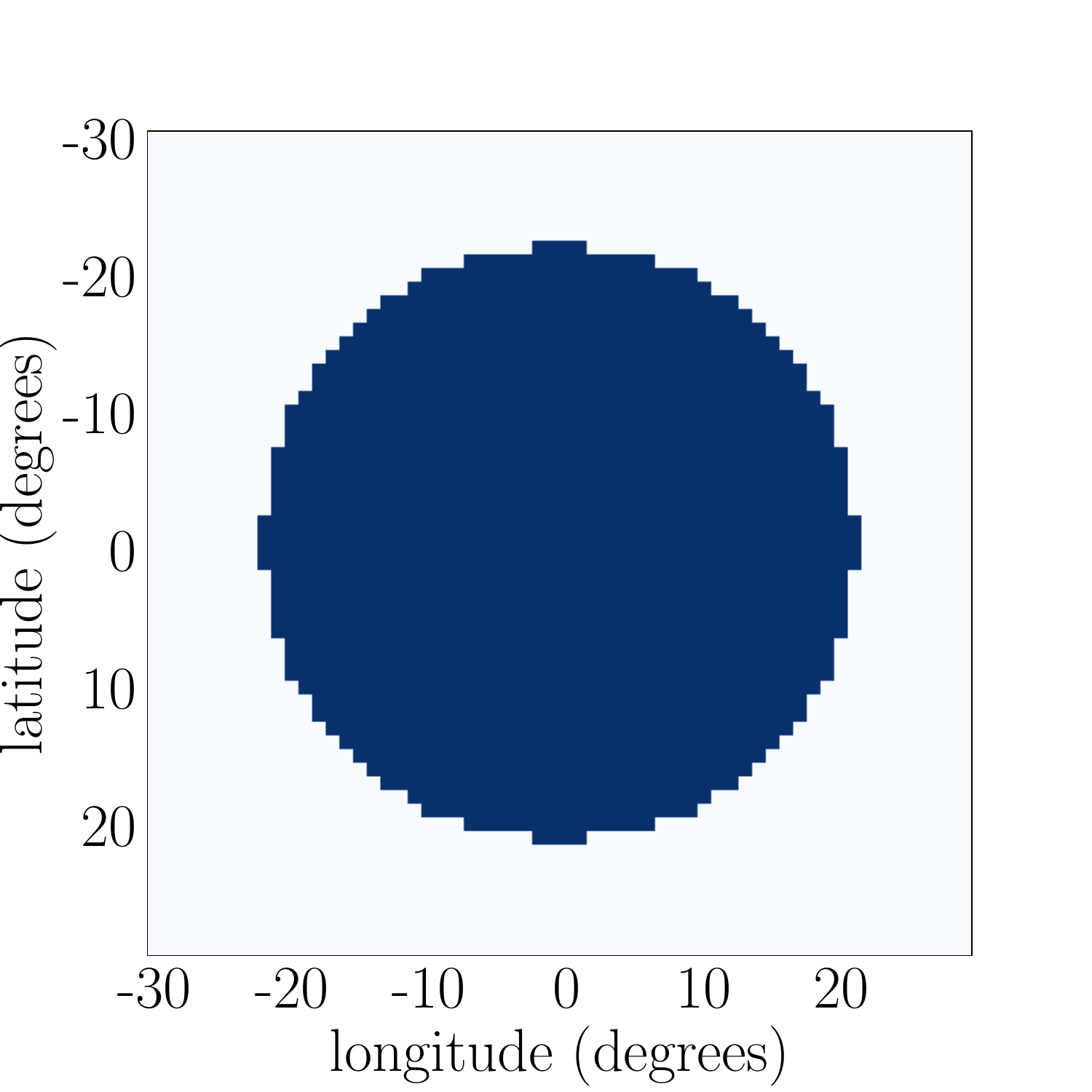}
\newline
\\
\centering
    Background Noise
\\
\includegraphics[width=0.32\linewidth]{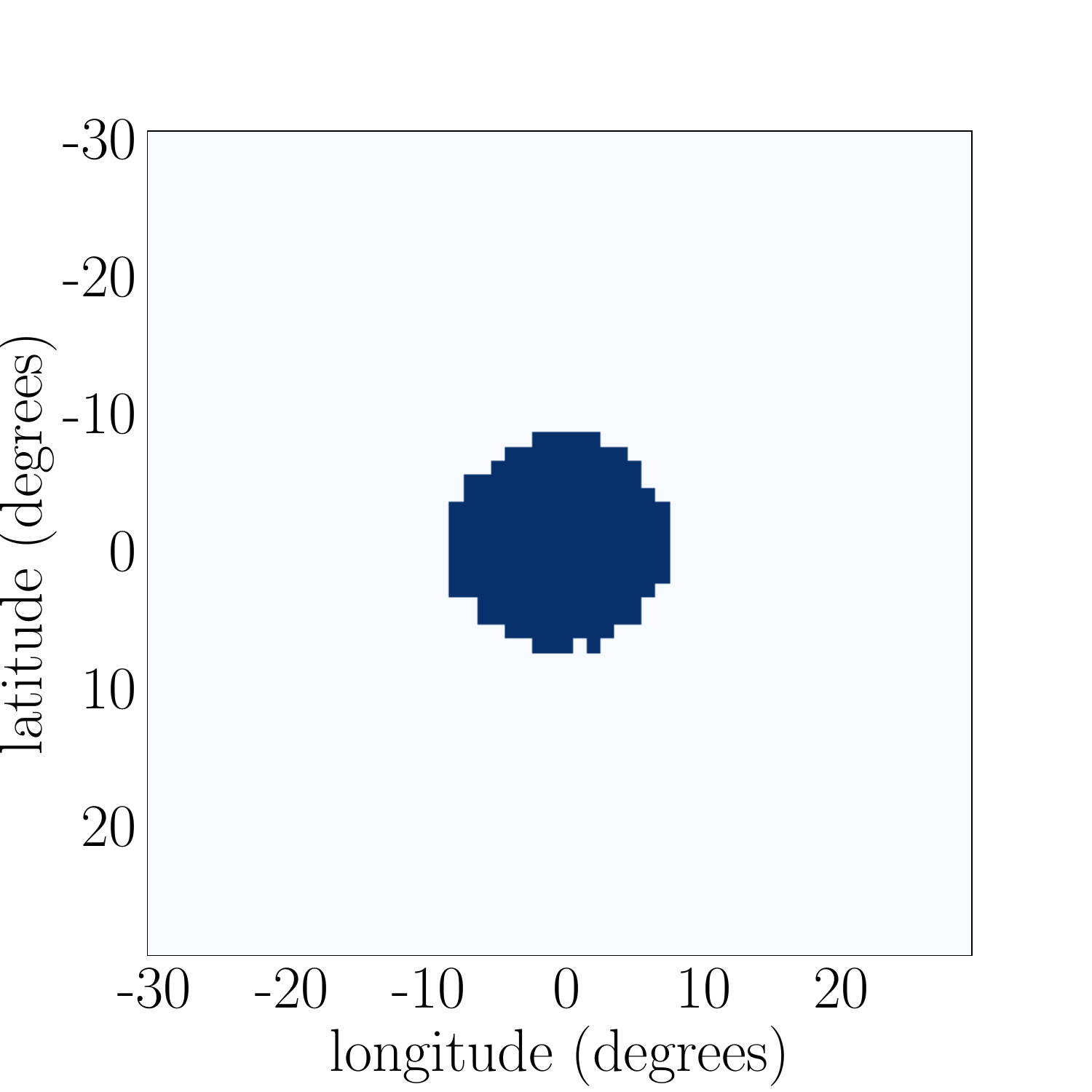}
\includegraphics[width=0.32\linewidth]{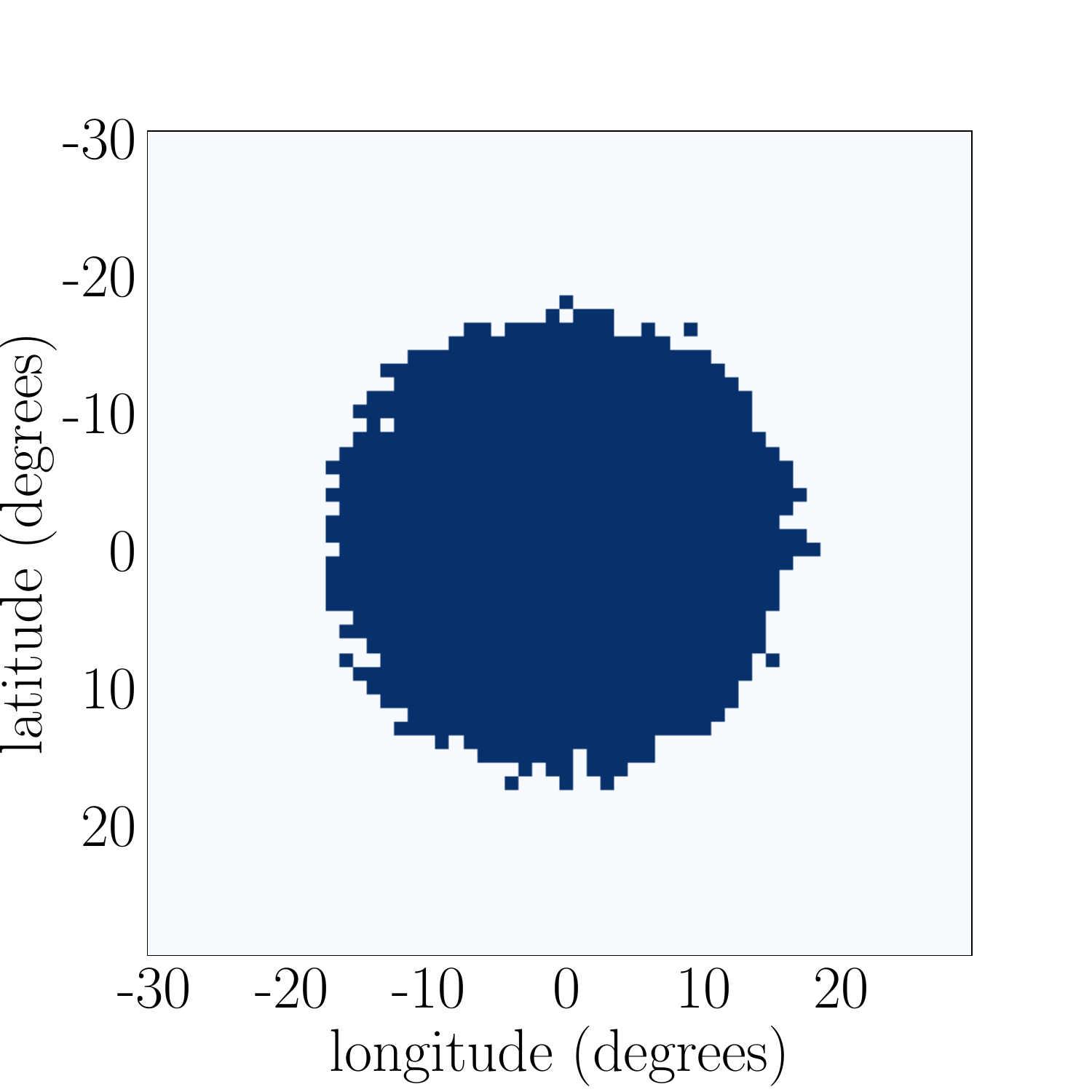}
\includegraphics[width=0.32\linewidth]{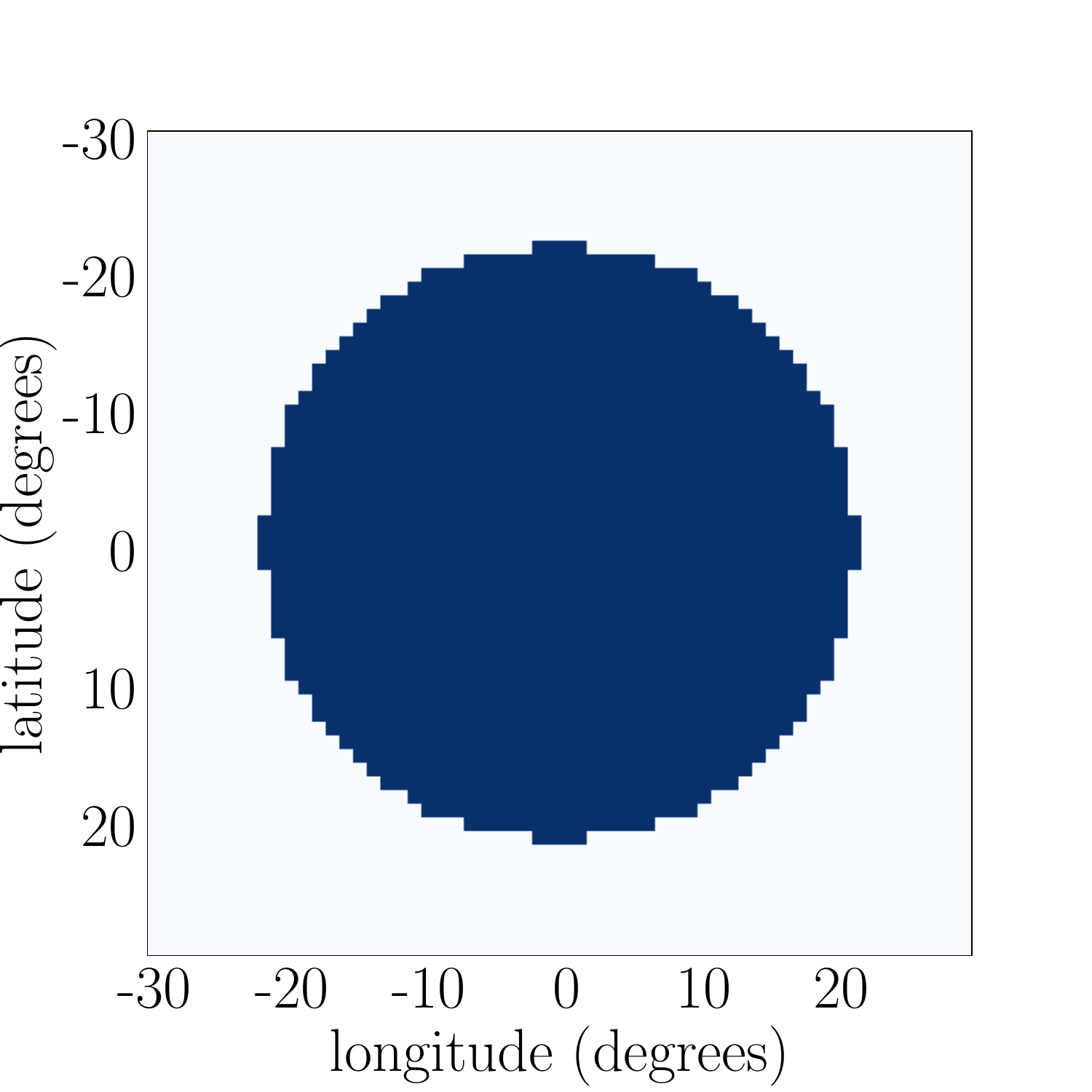}
\newline
\\
\centering
    Confidence Level
\\
\includegraphics[width=0.32\linewidth]{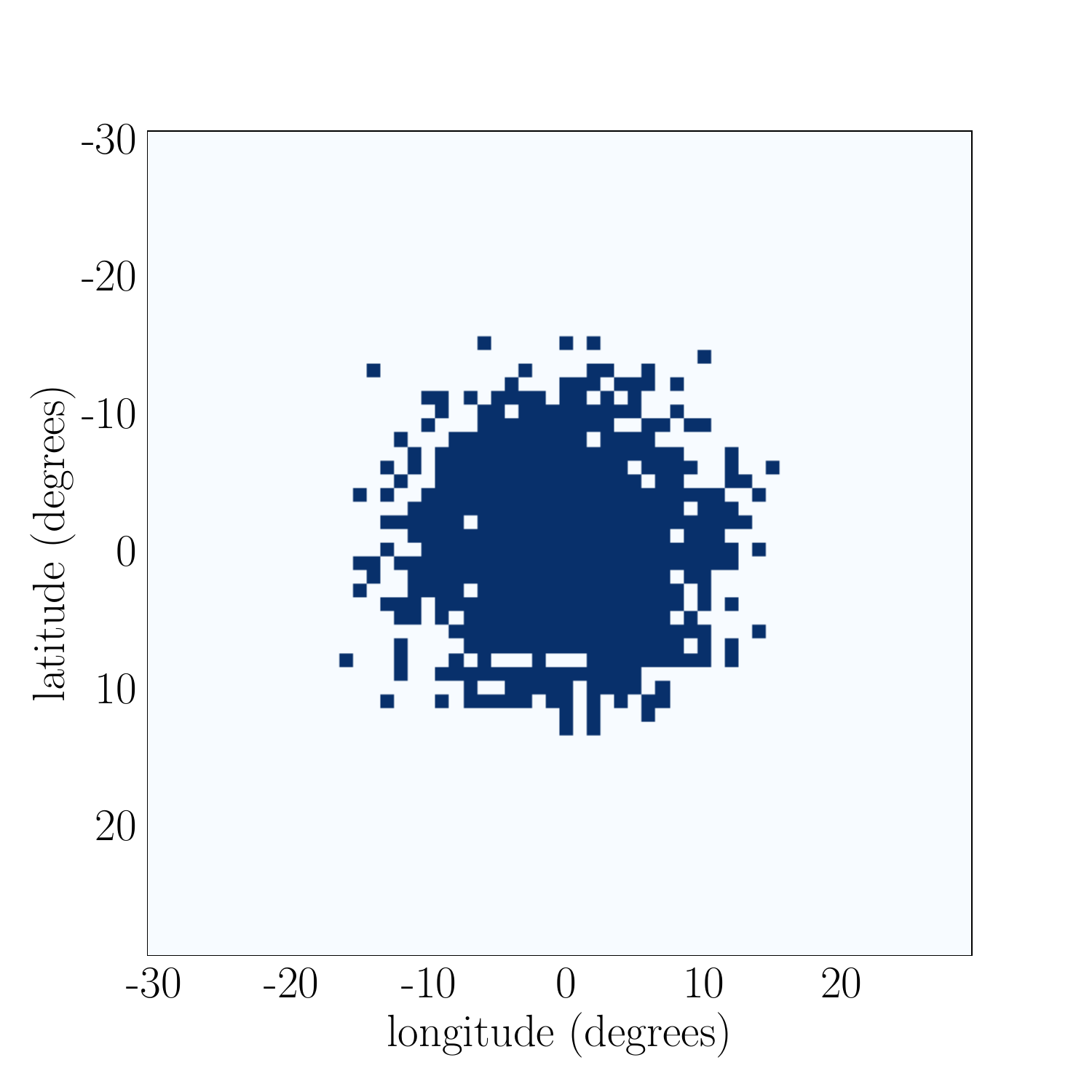}
\includegraphics[width=0.32\linewidth]{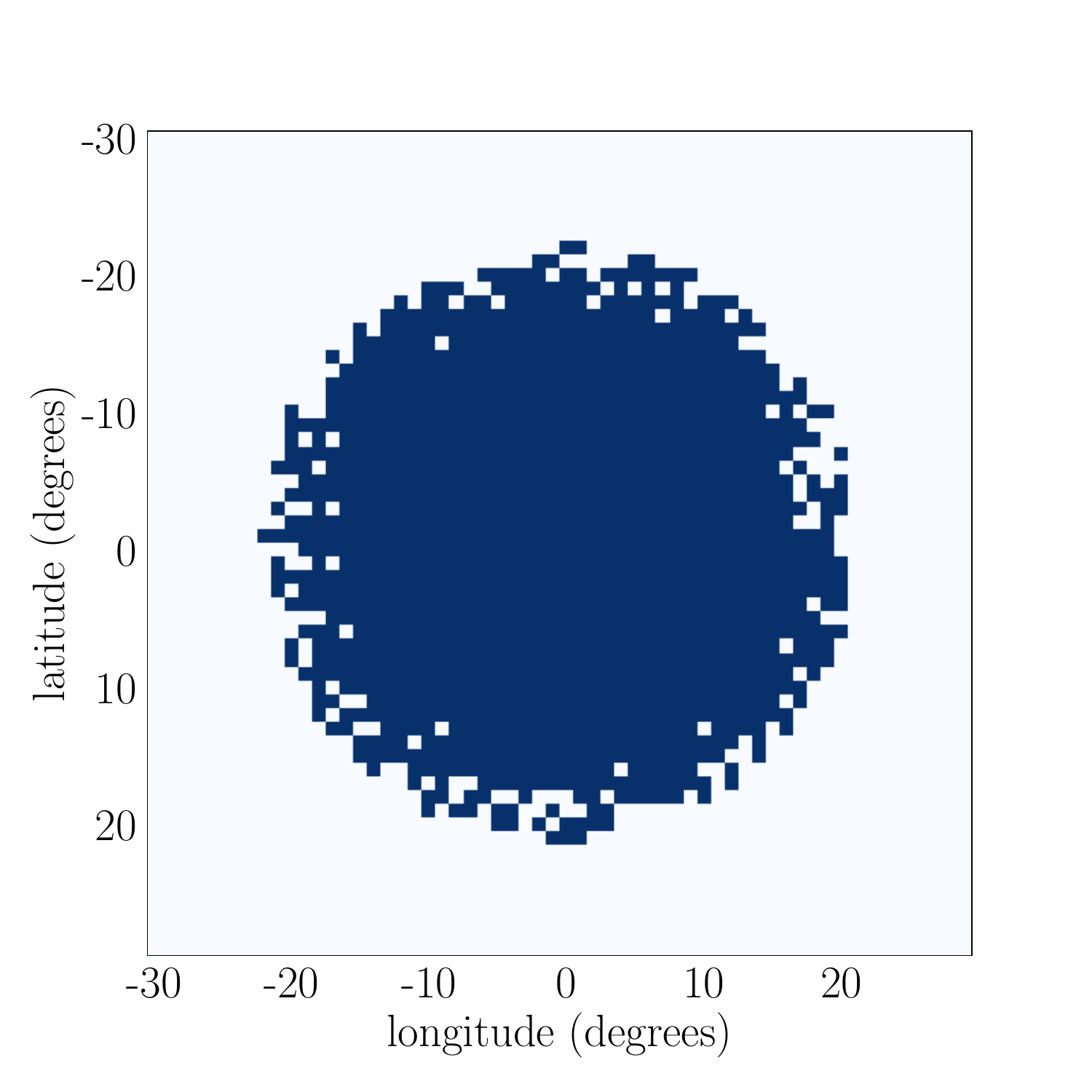}
\includegraphics[width=0.32\linewidth]{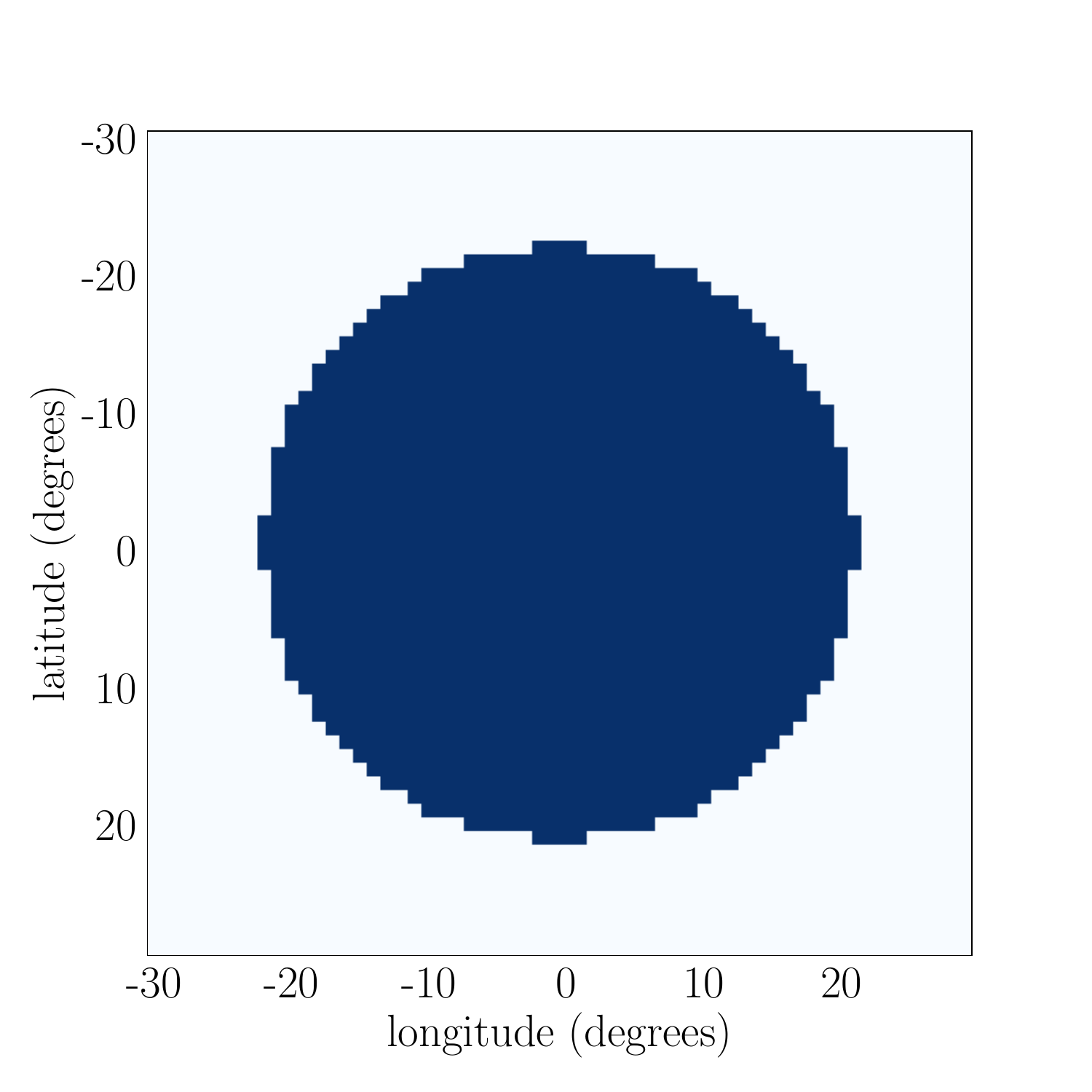}

\caption{
Bell violation shadows for a single downlink scenario. The satellite is in a polar orbit at 500 km altitude. The shadow is a snapshot when the satellite is just overhead the intersection of the Prime Meridian and the Equator. The dependence of shadow size on the acquisition time is shown on the top row. Confidence level is set to 1$\sigma$ and we consider 1 kHz background noise and 0.1 kHz dark count rate. The shadow size increases with increasing acquisition time, $t_{\rm acq} = 0.5 \rm ms$ (Top left) to $t_{\rm acq} = 1 \rm ms$ (Top middle) and finally for $t_{\rm acq} = 5 \rm ms$ (Top right) the shadow size saturates becoming equal to the visibility shadow. The dependence of shadow size on the background noise is shown in the middle row. The confidence level is set to 1 $\sigma$ with acquisition time  $t_{\rm acq} = 0.5 \rm ms$ and we consider a 0.1 kHz dark count rate. The shadow size increases with decreasing background noise, $r_{\rm bkg} = 50$ kHz (Middle left) to $r_{\rm bkg} = 10$ kHz (Center), and finally for $r_{\rm bkg} = 1$ kHz (Middle right) the shadow size saturates becoming equal to the visibility shadow. For comparison, the pair-rate (ebit generation rate at source) is 10 MHz. The dependence of shadow size on the confidence level is shown in the bottom row. The acquisition time is  $t_{\rm acq} = 0.5 \rm ms$ and we consider a 0.1 kHz dark count rate and a 1.0 kHz background noise rate. The shadow size increases with decreasing confidence level, 5-$\sigma$ (Bottom left) to 3-$\sigma$ (Bottom middle), and finally for 1-$\sigma$ (Bottom right) the shadow size saturates becoming equal to the visibility shadow.
}
\label{fig:3x3}
\end{figure*}

Next, we study the impact that the choice of acquisition time, the required confidence level for the Bell test success, background noise, and dark counts have on the size of the Bell shadows.  The results are collected in  Fig.~\ref{fig:3x3} and can be summarized as follows. The shadow grows in size and becomes less sparse as the acquisition time increases. This is consistent with the intuition that, if enough statistics are available, the Bell violation shadows become equal to the visibility shadows. Fig.~\ref{fig:3x3} (top) shows that, for a 1-$\sigma$ confidence level, the Bell shadow becomes equal to the visibility shadow for approximately 5 milliseconds of acquisition time. The trends in Fig. \ref{fig:3x3} (middle) again show that Bell violation shadows correctly quantify the regions of high quantum correlations, and the shadows shrink as noise and/or dark count rates increase, because these contribute with random (classical) correlations in the measurement outcomes. Finally, from Fig.~\ref{fig:3x3} (bottom) it is clear that demanding a higher confidence translates to a smaller shadow.

As was discussed in the previous sections, for the double downlink scenario, the Bell violation shadows must be defined with respect to some ground station. For simplicity, in this example we choose the first ground station to lie directly underneath the satellite at a given instant. The shadow then refers to a region on Earth that can successfully perform a Bell test with a station directly underneath the satellite, and hence is a quantifier of scale and quality of the network.

The effect of changing the acquisition time, noise and confidence levels in the double-downlink case is only quantitatively different from the single-link case. Hence,  in order to avoid repetition, we discuss  only the qualitatively different effect between the two scenarios. 

An important new effect appearing in the double-downlink scenario is that the two links are in general not identical; e.g. the level of background noise and losses is generically asymmetric.  For instance, consider the situations where one ground station is in daylight and the other is in the dark, or when one ground station has better detectors or noise mitigation tools. 

\begin{figure}[ht]

\includegraphics[width=0.45\linewidth]{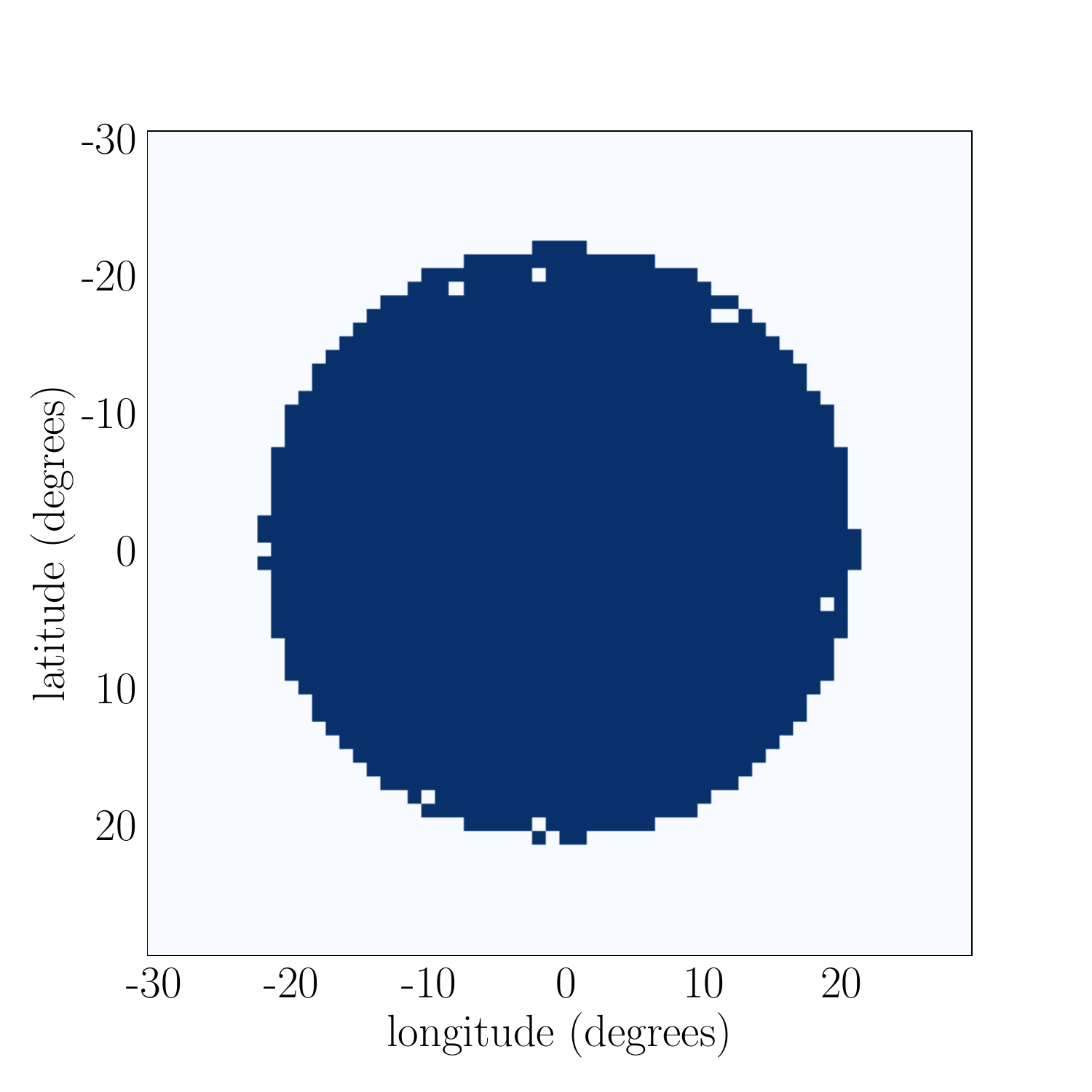}
\includegraphics[width=0.45\linewidth]{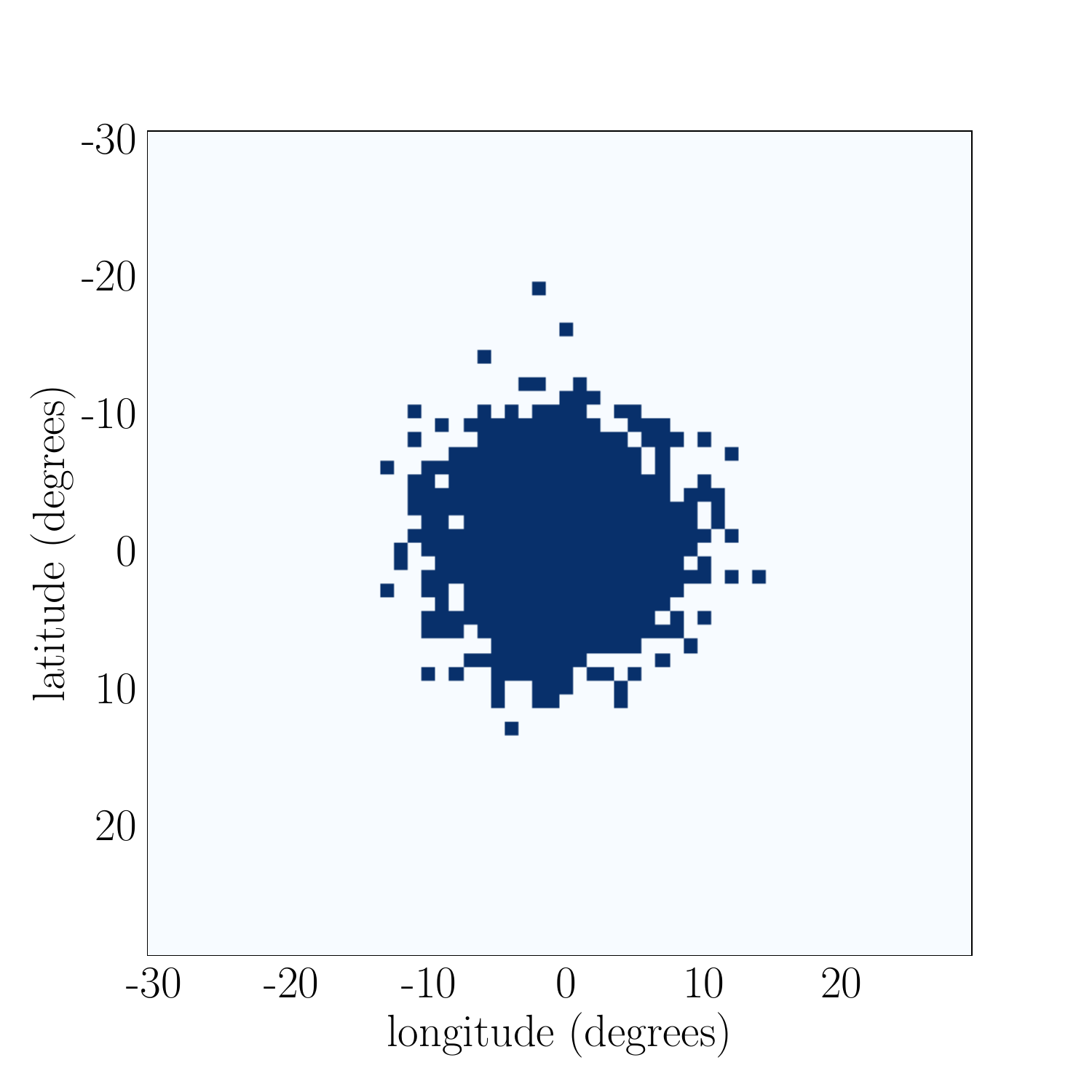}
\includegraphics[width=0.45\linewidth]{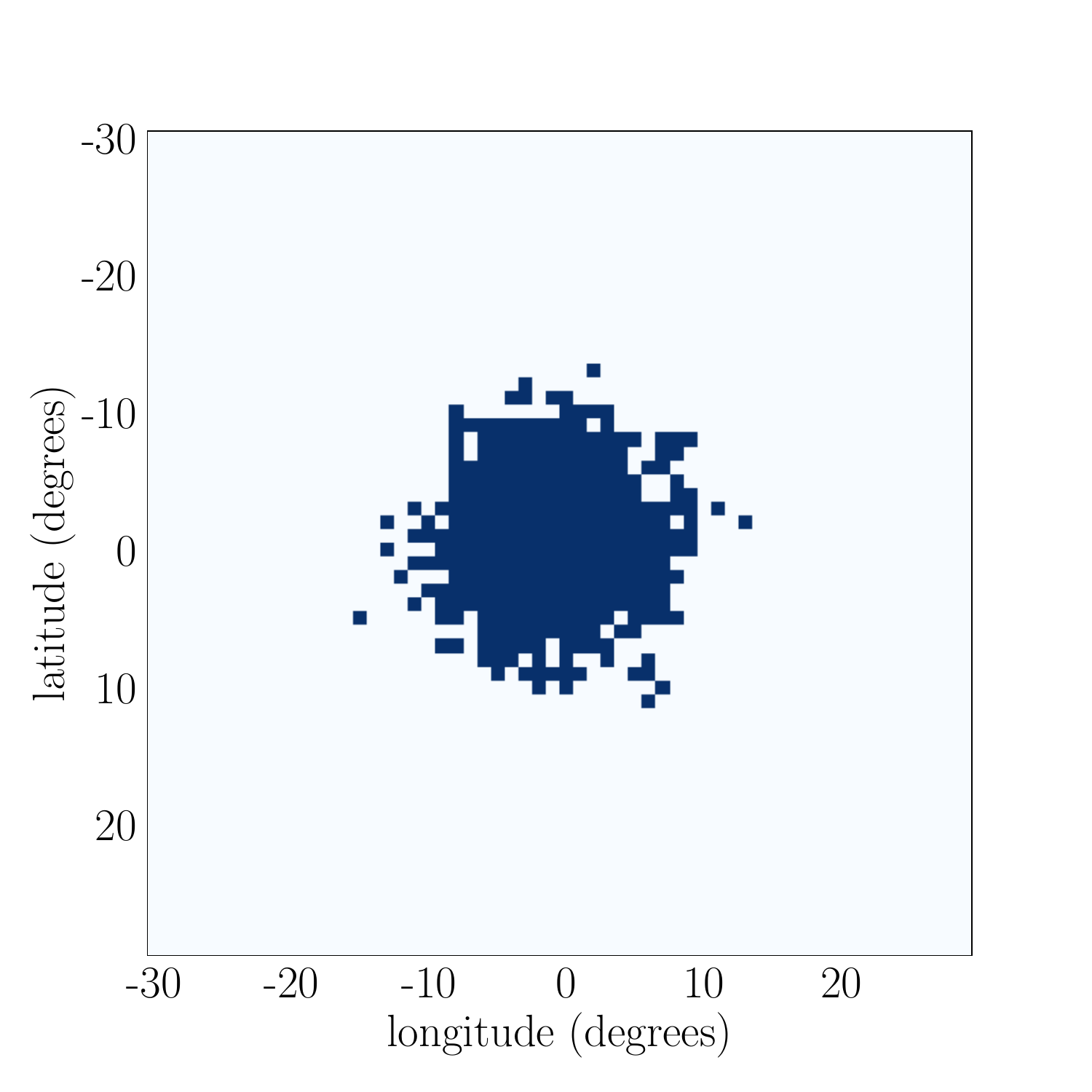}
\includegraphics[width=0.45\linewidth]{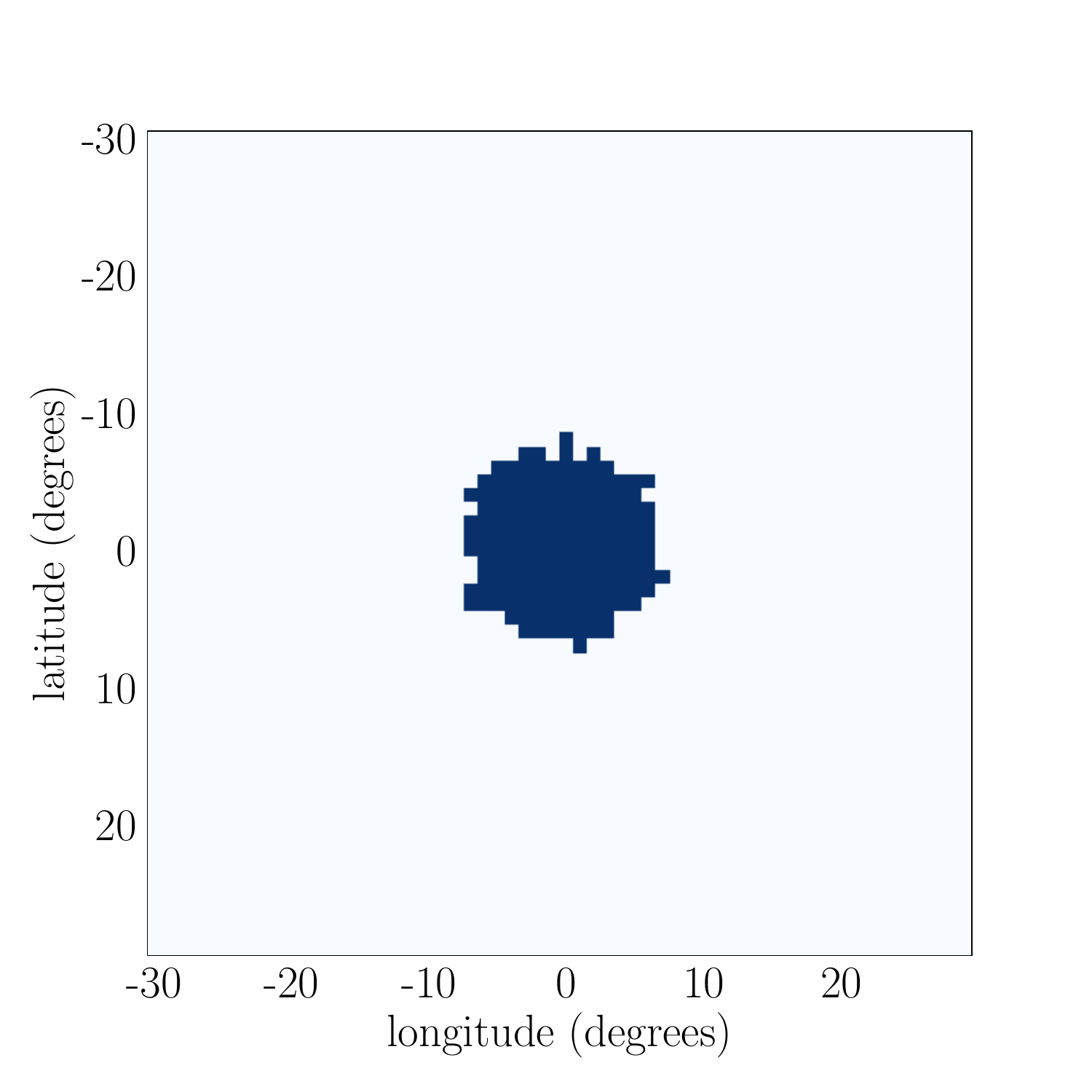}

\caption{Bell shadows for a double-downlink scenario. This figure shows the dependence of the size of shadow on the background noise rate (together with dark count rate). Comparison is made between situations with symmetric (left) and asymmetric noise levels (right) at the two ground stations. The geometric mean of noise levels is the same for the two compared cases. For the symmetric noise plots showed on the left, the noise levels in both arms are equal to 1kHz and 10 kHz for the top and bottom cases, respectively. For the corresponding figures on the right, the noise rates in the two channels are (100 Hz, 10 kHz) for the top figure, and (1 kHz, 100 kHz) for the bottom figure. In the asymmetric noise scenario, the larger noise rate dictates the shadow size and not the geometric mean of the two rates. In all plots, the acquisition time is set to 20 ms, the number of runs is 30, and we consider a 1-$\sigma$ confidence level. The satellite is in a polar orbit at 500 km altitude. The shadow is a snapshot when the satellite is just overhead the intersection of the Prime Meridian and the Equator. }
\label{fig:dd_asym_noise}
\end{figure}

One interesting message we extract from  our simulations is that it is the larger noise rate that dictates the shadow size, instead of, for example, the geometric mean of the two rates. This effect is shown in Fig.~\ref{fig:dd_asym_noise}.

The noisier channel dictates the shadow size by controlling the amount of quantum correlations that can be extracted from the arriving ebits. 
In the double-downlink scenario we can consider the background rate as effectively representing background plus dark count rate, which corresponds to the rate at which random photons are detected at the two ground stations independently. In contrast, for the single downlink scenario, dark counts have to be considered separately, since at the sender's (satellite) end only local detection takes place (i.e., the satellite does not need a telescope to receive background photons from the Earth). Consequently, dark counts are the dominant source of random correlations, since noise rates are very low.

\section{Protocols utilizing Bell violation shadows}
\label{sec:apps}
Having introduced and studied aspects of  Bell shadows in the previous sections, we move to show its versatility for different quantum applications. We will discuss the utilization of Bell violation shadows for situations where some classical information is shared, and quantum resources are used to add an extra level of security. Some examples are quantum key distribution protocols, quantum secured time distribution, and navigation schemes. We will also discuss quantum communication protocols, such as quantum networks. These applications will also cover the three scenarios, viz., single downlink (and/or uplink), double-downlink and connected satellites discussed in Section \ref{sec:Intro} (Fig.~\ref{fig:scenarios}).

\subsection{Quantum key distribution}
\label{subsec:QKD}
For quantum key distribution, Bell violation is an important test for quantum security. In particular,  the limitations introduced by finite statistics are important for the security of these protocols \cite{Cai2009}. As a direct test of this dependence, we look at a concrete QKD protocol, namely the entanglement based version of BB84 protocol (or Ekert 91) \cite{Lo1999, E91}. The quantity of interest that ensures the success and security of this protocol is, like for most other QKD protocols,  the quantum bit error rate (QBER) \cite{RevModPhys.94.025008}. The QBER is defined as the fraction of measurement outcomes for which the key shared between Alice and Bob differs from each other. The larger the QBER, the greater the chances of successful eavesdropping and the lower the security of the key. Moreover, quantitatively, for the asymptotic regime (infinite statistics) and symmetric attacks, it has been shown that a QBER $< 11 \%$ ensures security for entanglement-based BB84 protocol utilizing two measurement settings for each qubit \cite{RevModPhys.94.025008}. 
For our simulations, we consider two measurement settings for key generation and two for the Bell test. The Bell test bases are chosen optimally (to get a maximal violation for a Bell state), i.e., $\{\Pi_{0^\circ}, \Pi_{45^\circ}\}$ and $\{\Pi_{22.5^\circ}, \Pi_{-22.5^\circ}\}$ for Alice and Bob, respectively. For the generation of the key, we use the $\{\Pi_{0^\circ}, \Pi_{45^\circ}\}$ bases at both Alice's and Bob's ends. 

\begin{figure*}[ht]
    \centering
    \includegraphics[width=\textwidth]{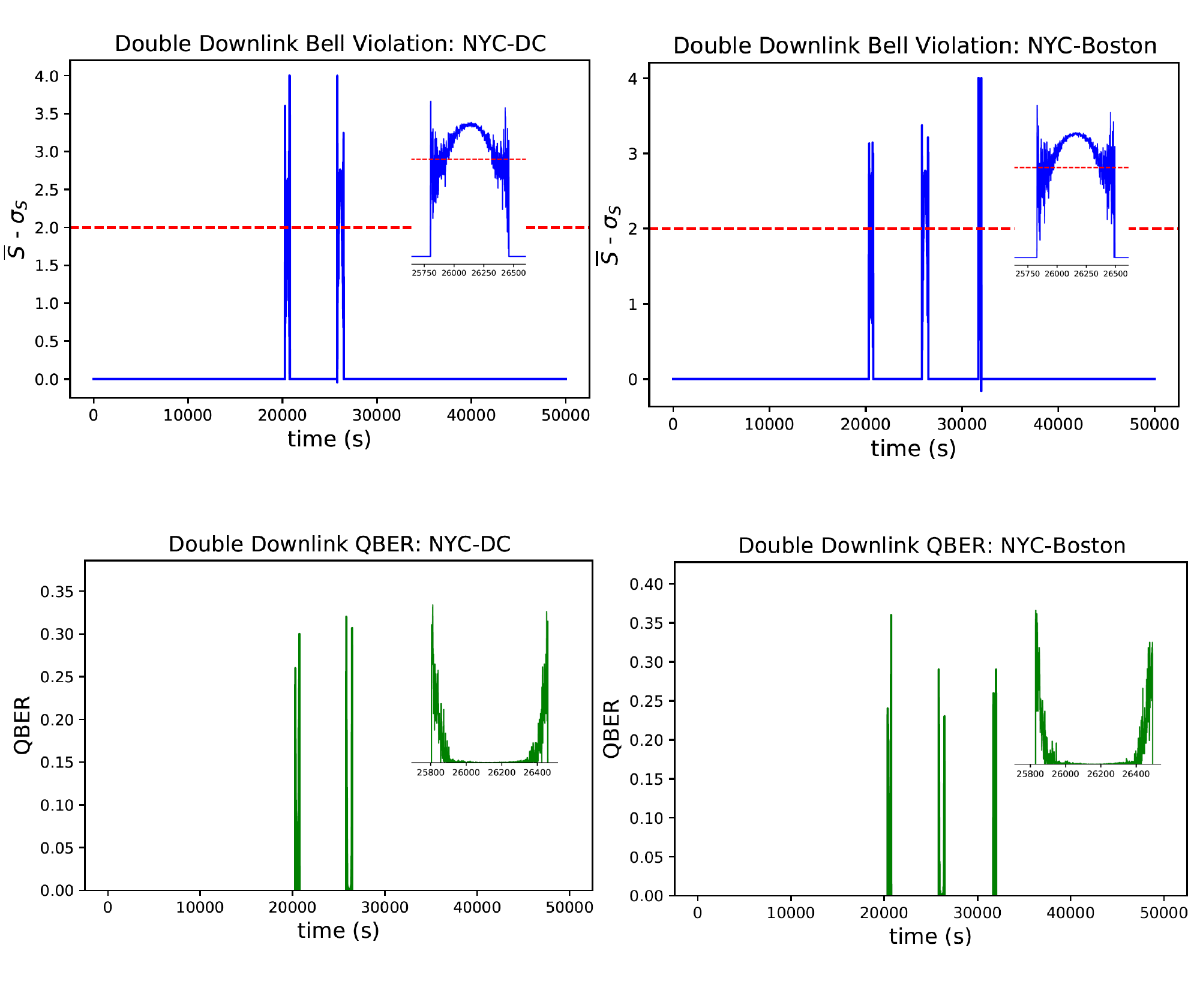}
    \caption{The top two figures represent the average CHSH value minus the standard deviation over the period of 50,000s for the double downlink scenario with ground stations at  New York City and DC (left), and New York City and Boston (right), using a single satellite in a polar orbit (500 km altitude) along the Prime Meridian. The zoomed in insets represent the rightmost peak on the left figure and the middle peak on right figure. A red dashed line is drawn at $\Bar{S}-\sigma_{S}$ = 2.0 to indicate that above this line, the Bell test can be successful with 1-$\sigma$ confidence. For comparison, in the bottom two figures we plot the QBER (quantum bit error rate) for the same pairs of cities, with the zoomed-in sectors coming from the same peaks as the CHSH figures. For periods during  which the satellite is not simultaneously visible to both pairs of cities, the QBER is not defined, and thus not included on the plot. Note that a lower QBER value represents the ability to generate a more secure quantum key. At first sight, and due to the small width of the visibility peaks (approximately 700 s) compared to the total simulation time scale (more than 12 hours), it might seem that the peaks of the CHSH number and QBER value occur simultaneously. This is not the case. To clarify this and  highlight the relation between QBER and Bell violation trends, we refer the reader to the inset figures. In the insets one can clearly observe that the maxima of the two upper Bell violation plots coincide exactly with the minima of the lower two QBER plots (within the same visibility period). Also larger the fluctuations in CHSH number the larger the QBER value is (see the tails of both the inset plots.) Finally, since NYC is slightly closer to Boston than to DC, in the simulated time period NYC and DC can share quantum key twice, but NYC and Boston can share key thrice. In Fig.~\ref{fig:QBER_vs_Bell} we will show the scale of this QKD network.}
    \label{fig:NYC_qkd_network}
\end{figure*}

Consider the double-downlink scenario with the two ground stations trying to establish a quantum-secure key, with the satellite playing the role of the entanglement source. The satellite is still assumed to be in a polar LEO at 500 km altitude. We simulate a QKD network over the continental United States using a single satellite. In order to share ebits, two cities in the network must be simultaneously visible to the satellite, thus putting constraints on the size of the network. Also, as discussed in Section \ref{sec:Bell_shadow}, the shadow here is defined by fixing one of the ground stations. In Fig.~\ref{fig:NYC_qkd_network} we choose this to be New York City (40.7128° N, 74.0060° W). We show the time trace of $\bar{S}-\sigma_S$ for city pairs New York City -- Washington D.C. (38.9072° N, 77.0369° W) and New York City -- Boston (42.3601° N, 71.0589° W). This quantity being $\geq 2$ indicates Bell violation at a 1-$\sigma$ confidence level. We see two to three peaks during a 1-day simulation. 

Furthermore, as shown in the inset of Fig.~\ref{fig:NYC_qkd_network}, close to the time when the satellites just reach over (or go below) the horizon for both the cities, the value of the Bell violation fluctuates. This is because the ebit-rate is non-zero but low compared to the background noise rate. In terms of the Bell violation shadow, this occurs when the two cities are close to the edges of the shadow. On the other hand, when the satellite is closest to both ground stations and a large number of ebits are received, the value of the CHSH number is stable, and manifested in an even higher confidence Bell violation. 
In  Fig.~\ref{fig:NYC_qkd_network} (bottom) we show the corresponding QBER time trace. It is clear that the QBER rates are low ($<11\%$) and non-fluctuating whenever the Bell violation is high and non-fluctuating, that is, when the two cities are closest to the satellite (near the shadow center).

\begin{figure}[h]
    \centering
    \includegraphics[width=\linewidth]
    {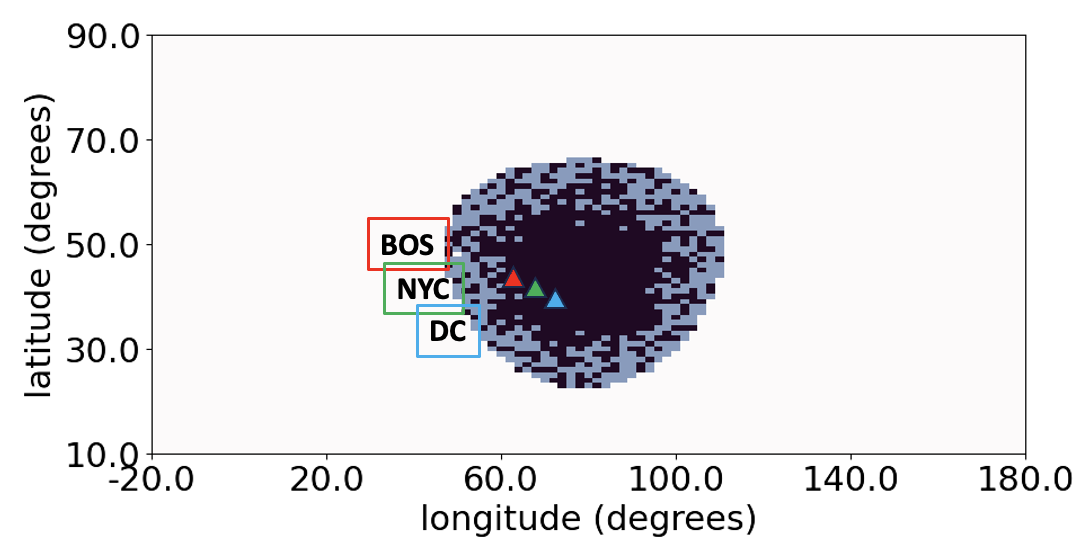}
    \includegraphics[width=\linewidth]
    {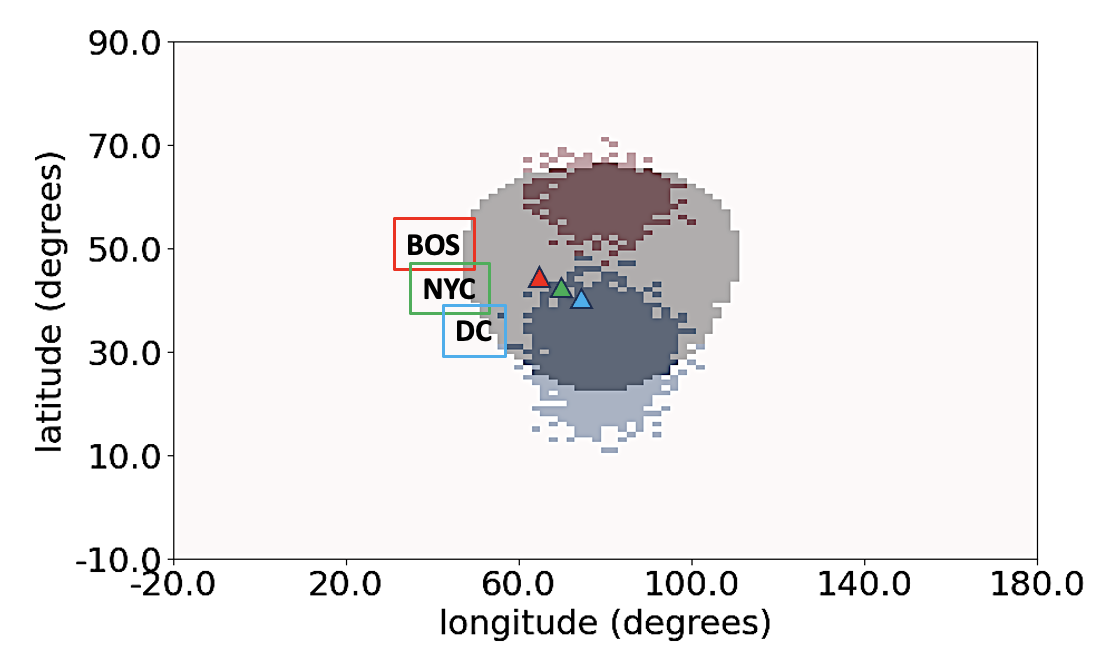}
    \caption{(Top) Overlap of QBER (light blue) and Bell violation shadows (dark blue) for a double-downlink scenario using NYC as the reference city, when the satellite passes close to NYC. The QBER shadow defines the region on Earth with which NYC can perform QKD with a QBER of less than $11\%$. Considerable overlap between the two shadows underscores the usefulness of Bell violation shadows for assessing QKD network scale and quality. (Bottom) Three snap-shots of the QBER shadows, corresponding to three different instants for the same scenario. The total time elapsed in this shadow track is 450 seconds, which is close to the visibility period from NYC. Notice that, as the shadow moves away from NYC, the QBER shadow becomes smaller. Also for the red (top-most) shadow, NYC is in fact outside the shadow. We explain this effect in the main text. The satellite is in a polar orbit at 500 km altitude and background noise rate is $10$ kHz.}
    \label{fig:QBER_vs_Bell}
\end{figure}

Next, we address the question of network scale, i.e., what are the farthest cities from NYC that can share quantum keys with it, using the single LEO-satellite configuration discussed above (this is the same configuration discussed in  Fig.~\ref{fig:dd_shadow}). 
In Fig.~\ref{fig:QBER_vs_Bell} we plot two shadows; one is the Bell violation shadow and the other is the QBER shadow. The QBER shadow is defined as the region on Earth where two ground stations can share a key using an entanglement based BB84 protocol with a QBER of less than 11\%. It is clear that Bell violation shadows (dark blue-dark grey) show considerable overlap with the QBER shadows (light blue-light grey).

Furthermore, as the satellite moves, the cities move through different regions of the Bell shadow (dense or sparse) which provides quantitative information about the QBER. Sparse regions in Bell shadow have a highly fluctuating and high QBER, whereas dense regions have a low and stable value of QBER, meaning they are suitable for a successful QKD protocol. 

We re-emphasize an aspect already mentioned in the previous section: it is possible that, at certain instants, a city can be outside the Bell or QBER shadow defined with respect to itself. In Figure \ref{fig:QBER_vs_Bell}, (bottom) we see that, for the red (top-most) shadow, NYC, Boston and DC are all outside the shadow. This means that NYC cannot perform a successful QKD protocol with cities immediately in its vicinity (Boston and DC), but it can do so with cities further apart, which are within the shadow, because these cities are closer to the satellite. This is an interesting feature of the double-downlink scenario. In general, as long as a ground station is visible from the satellite, there might exist a region on Earth directly underneath the satellite that can conduct a statistically significant Bell test or QKD protocol with the fixed ground station.

It is important to note that, for finite statistics, the QBER $< 11 \%$ condition does not hold exactly, and corrections must be made. Such corrections, referred to as finite key analysis, have been thoroughly studied in the literature \cite{Sidhu2022} and are not the focus of this work. We intend to simply show a strong correlation between Bell violation shadows and their QBER counterparts, and thus highlight the use of the former as a good quantifier of network scale and quality for entanglement based protocols.
 
\subsection{Quantum clock synchronization}
\label{subsec:qcs}
The next protocol we study is a satellite-based protocol to synchronize clocks on Earth using quantum resources, first introduced in \cite{Lee2019,AntiaTroupeQCS,paper1} and extended in \cite{paper2,paper3}. This is a useful protocol not only from the point of view of being a high-precision time distribution scheme with a layer of quantum security, but also because it is a precursor for a global quantum GNSS system \cite{white_paper,paper3}. Additionally, quantum clock synchronization protocols (QCS) share many features with a satellite-based Quantum Internet \cite{LSU_satsim}.  

Very briefly, the protocol is described as follows (for details, see Refs.~\cite{paper1,paper2,paper3}).
 Both the ground station  (Alice) and satellite (Bob) are equipped with entangled photon-pair sources. When a ground station becomes visible to the satellite, both the ground station and satellite share entanglement pairs to synchronize their clocks utilizing the tight time-of-birth correlations (tens of femtoseconds \cite{Shih2004}) of entangled photons generated via an SPDC source. The relative velocity between the satellite and ground station also affect the precision of synchronization. As the satellite moves. it can synchronize with different ground stations, in turn, synchronizing them with each other. At the single link level (one ground station synchronizing with one satellite), which is a two-way link between the ground station and satellite, the sync precision, which we denote as $t_{\rm bin}$, is given by the following equation \cite{paper2}:
\begin{eqnarray}
    t_{\rm bin} = \frac{N_{\rm min} v_{\rm rel}^{\rm rad}}{R \eta c},
    \label{eqn:t_bin}
\end{eqnarray}
where $c$ is the speed of light in vacuum, $R$ is entangled pair generation rate at the source, $v_{\rm rel}^{\rm rad}$ is the relative radial velocity, and $\eta$ is transmissivity of the quantum communication channel between the ground station and the satellite. The former two are constants, while $v_{\rm rel}^{\rm rad}$ and $\eta$ the  are dependent on the relative positions of the two parties and change in time as the satellite move. $N_{\rm min}$ is defined as the minimum mean number of exchanged ebit pairs between the satellite and the ground station for the protocol to be considered successful.  
Since photon detection through a lossy channel is a Poisson process, the probability of success of the protocol in individual cases is determined by the mean $N_{\rm min}$. Further, as is clear from Eqn.~\eqref{eqn:t_bin}, $t_{\rm bin}$  is also proportional $N_{\rm min}$. Thus, the higher the demand on the success probability of the synchronization protocol, the worse the precision of the sync (i.e., larger $t_{\rm bin}$).

Using tools developed in a previous work \cite{paper2}, and using Eqn.~\eqref{eqn:t_bin}, we evaluate in this article the sync precision shadows of satellites. Precision shadows are regions on Earth where a satellite can provide time synchronization at a certain minimum precision between clientele within the shadow. Therefore, the precision shadow quantifies the scale and quality of the QCS protocol. 

\begin{figure}[h]
    \centering
    \includegraphics[width=\linewidth]{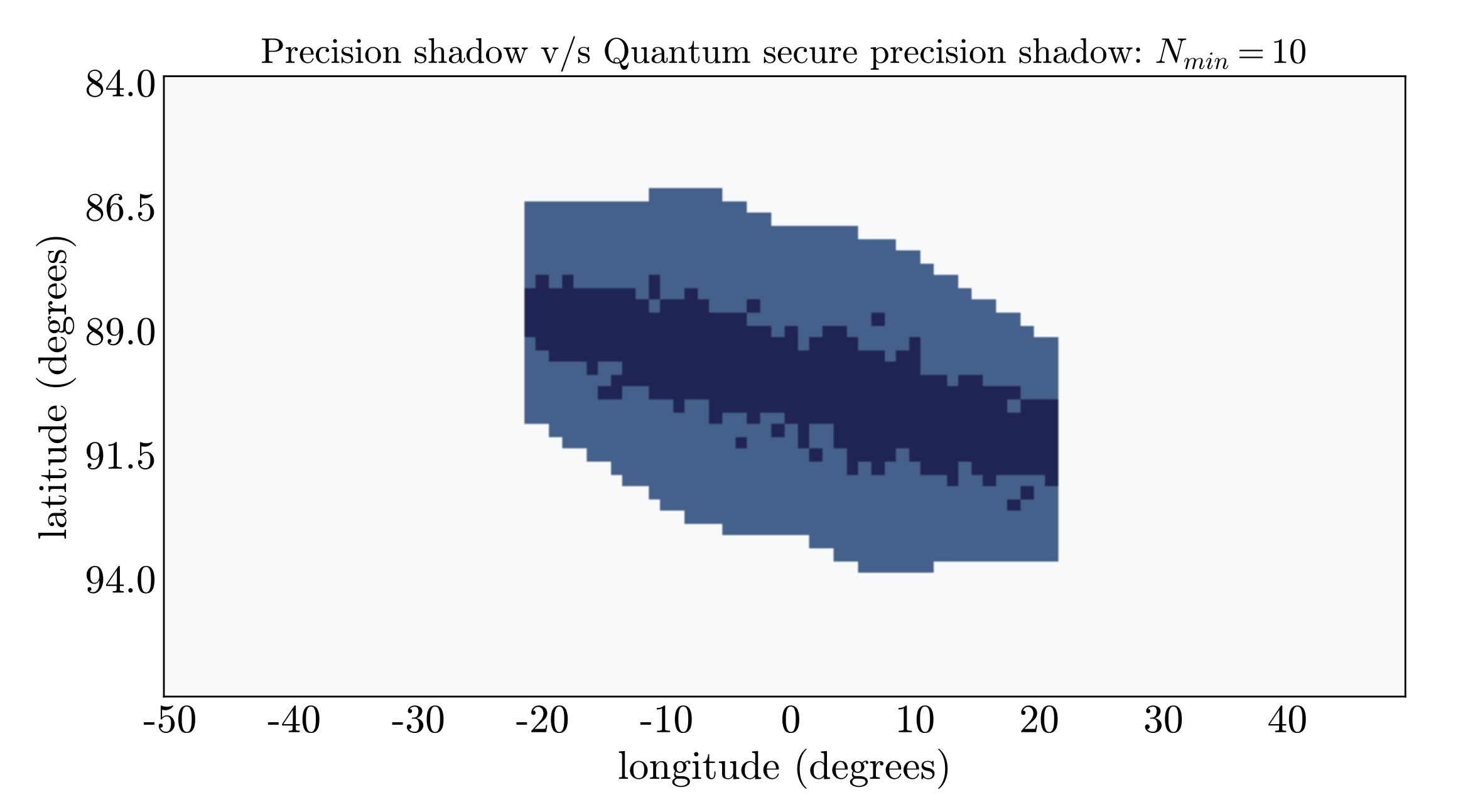}
    \includegraphics[width=\linewidth]{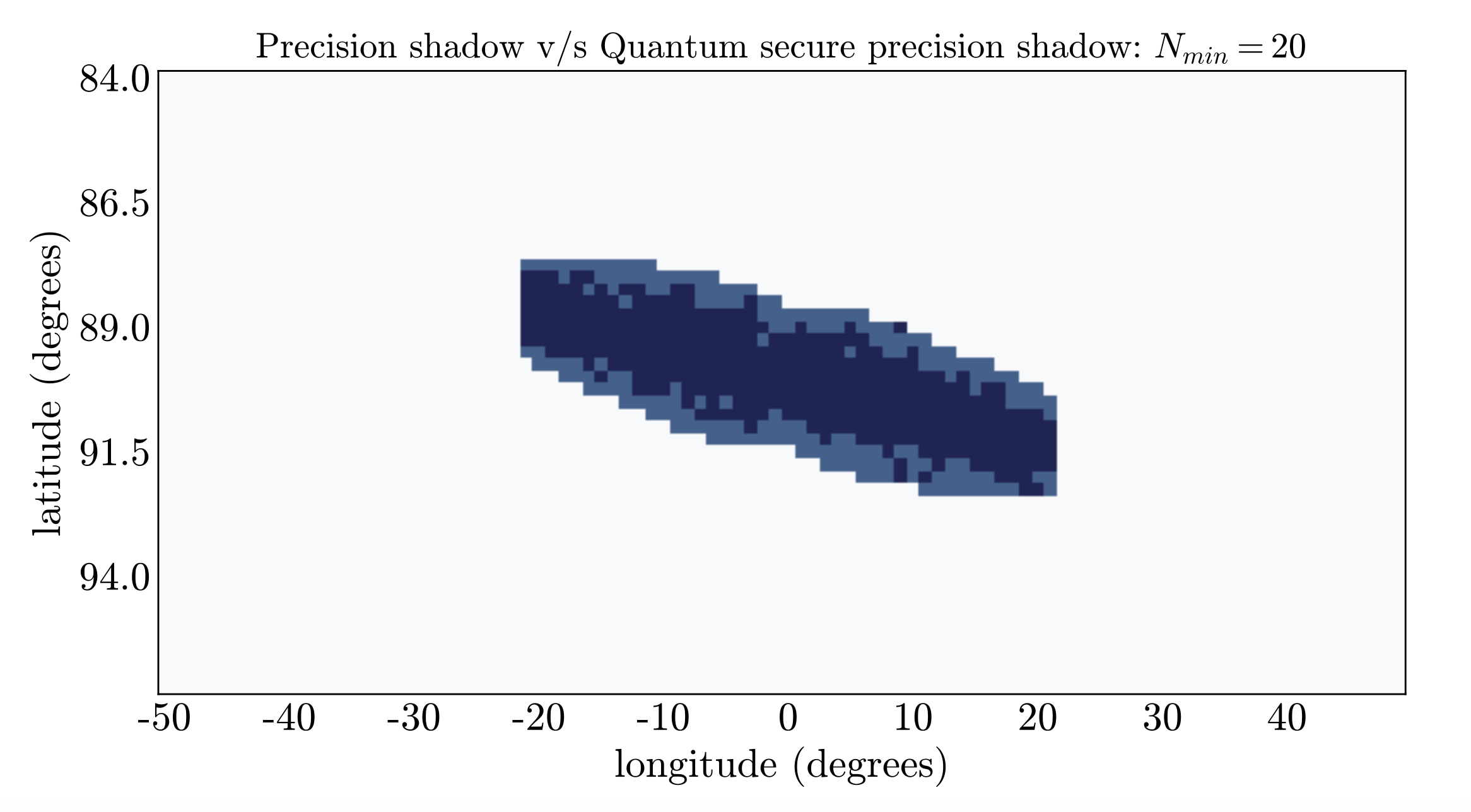}
    \includegraphics[width=\linewidth]{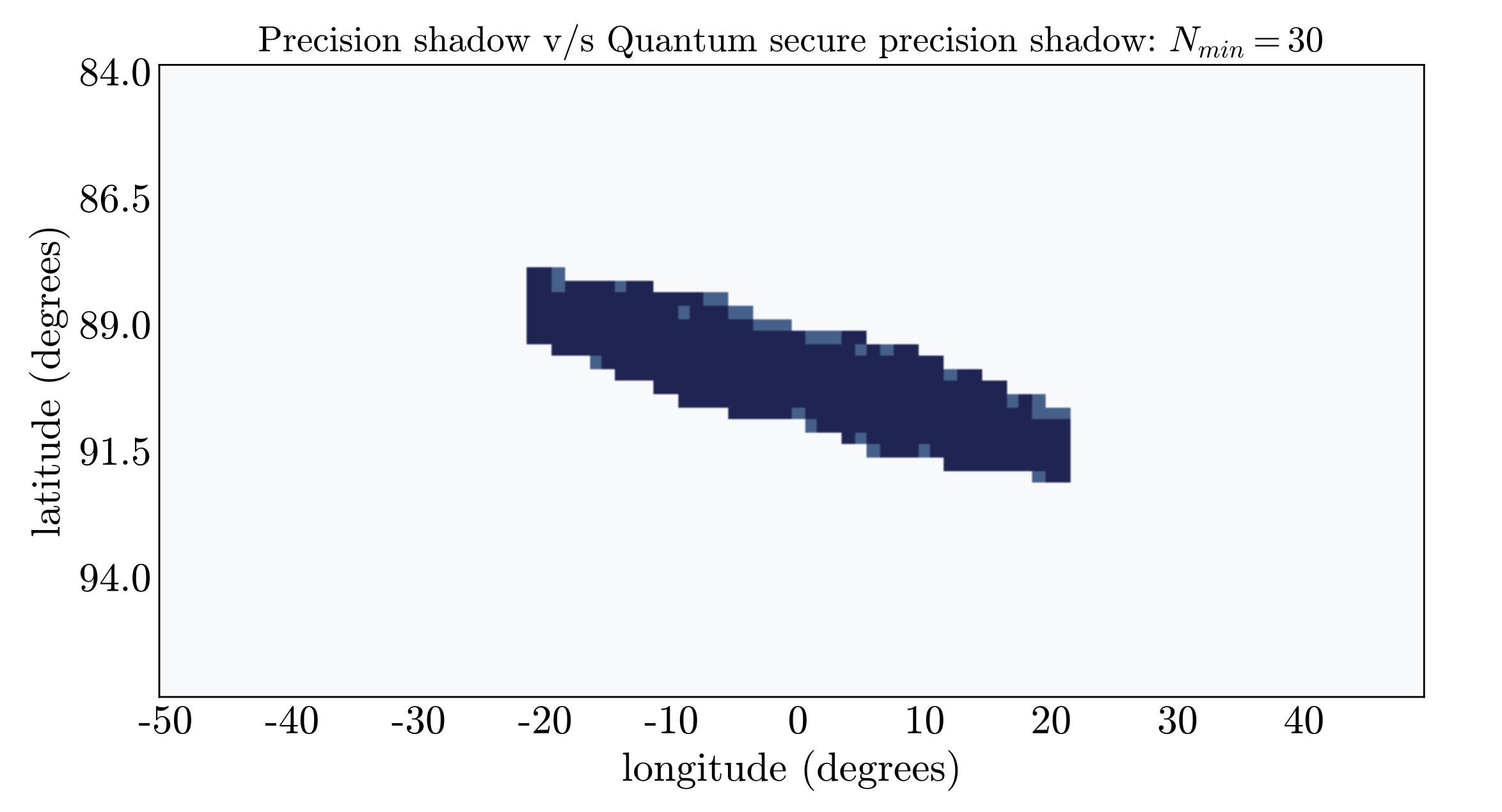}
    \caption{1ns sync precision shadows with (dark blue) and without (light blue) the constraint that Bell inequality be also violated at 1-$\sigma$ confidence, for different values of $N_{min}$. $N_{min} = 30$, ensures that quantum security can also be verified whenever the QCS protocol is successful.}
    \label{fig:secure_precision_shadow}
\end{figure}

In Fig.~\ref{fig:secure_precision_shadow}, we compare the precision shadows for 1~ns time synchronization and a ``quantum secure precision shadow'', which we define as the intersection of the precision shadow with the single-uplink Bell shadow computed at the 1-$\sigma$ confidence level. We use the uplink since it is the weaker (lower shared ebit rates) of the two QCS links. For this protocol, it makes more sense to compare the precision shadow with the quantum secure precision shadow, instead of the Bell shadow itself, because the latter is not sensitive to the asymmetries introduced by relative radial velocities. 
We see that, as $N_{\rm min}$ is increased (from top to bottom), the precision shadow starts to shrink and eventually becomes identical to its quantum-secure counterpart.
This shows a clear relationship between the success probability of achieving a certain level of precision (corresponding to a value of $N_{\rm min}$) and achieving quantum security via Bell tests ---two independent aspects of the clock synchronization protocol that can nevertheless be viewed through the common lens of Bell shadows developed in this work. Bell shadows provide a new operational meaning to the choice of $N_{\rm min}$; it needs to be chosen such that it ensures a successful 1-sigma Bell test. Of course, as the value of $N_{\rm min}$ is increased the precision shadow shrinks, since demanding a higher success probability for QCS protocol shrinks the region in which it can be achieved. At the same time this also ensures a high confidence Bell test to be successful between the satellite and any ground station in this new smaller shadow. This feature is another example of how our framework could be useful to other protocols.

\subsection{Quantum networks}
\label{subsec:Quantum_Networks}
The development of large scale and robust quantum networks will play a pivotal role in the implementation of several quantum technologies, including distributed quantum computing, distributed sensing and long distance QKD etc. \cite{Wehner2018,van2014quantum,Simon2017, dowling2020schrödinger}. Since long distance entanglement distribution is a basic element of quantum networking protocols, satellite-based entanglement distribution will play an important role for the implementation of such networks \cite{Kozlowski2019}. In this section, we motivate how the concept of Bell shadows developed in this article can help inform the choice of particular implementations or designs of networking protocols.

More concretely, to obtain robust high-fidelity entanglement between two distant nodes, several imperfect (low fidelity states) ``quantum links'' created by each ebit pair must be distilled. But since photon pairs arrive at the nodes non-simultaneously, the use of quantum memories is usually proposed to store these elementary entangled links which are then distilled into a few high-fidelity states via entanglement distillation \cite{memory_review, Gndoan2021, Mol2023}. However, quantum memories have a finite coherence time after which the states stored in them become practically useless \cite{memory_review}. Large Bell violation indicates a high level of quantum correlations or entanglement between parties which corresponds to high-quality links between them. Furthermore, quantitative knowledge of the amount of quantum correlations is crucial from the point of view of entanglement distillation attempts, since the probability of success of distillation and the entanglement of the distilled state, both depend on the amount of entanglement in the input states \cite{BBPSSW96, Bennett_mixed}.

Keeping these relationships in mind, the concept of Bell violation shadow for a given acquisition time gets an operational meaning in the context of quantum networks. Both in the double-downlink and single-downlink scenario, the use of quantum memories will restrict the acquisition time for one round of entanglement distribution and/or verification. In the double-downlink scenario, the acquisition time can be directly interpreted as the coherence time of quantum memories on the two ground stations. Thus, the size of the Bell shadows quantify how much the network size ---the longest entanglement link--- increases when the coherence time of quantum memories is increased. At the same time, enforcing a higher confidence level ($n\sigma$ level) for the shadows would imply a higher success probability of generating a high-fidelity entangled state between two ground stations within the shadow. For example, see Fig.~\ref{fig:3x3} (top and bottom respectively).
\begin{figure}
    \centering
\includegraphics[width=\linewidth]{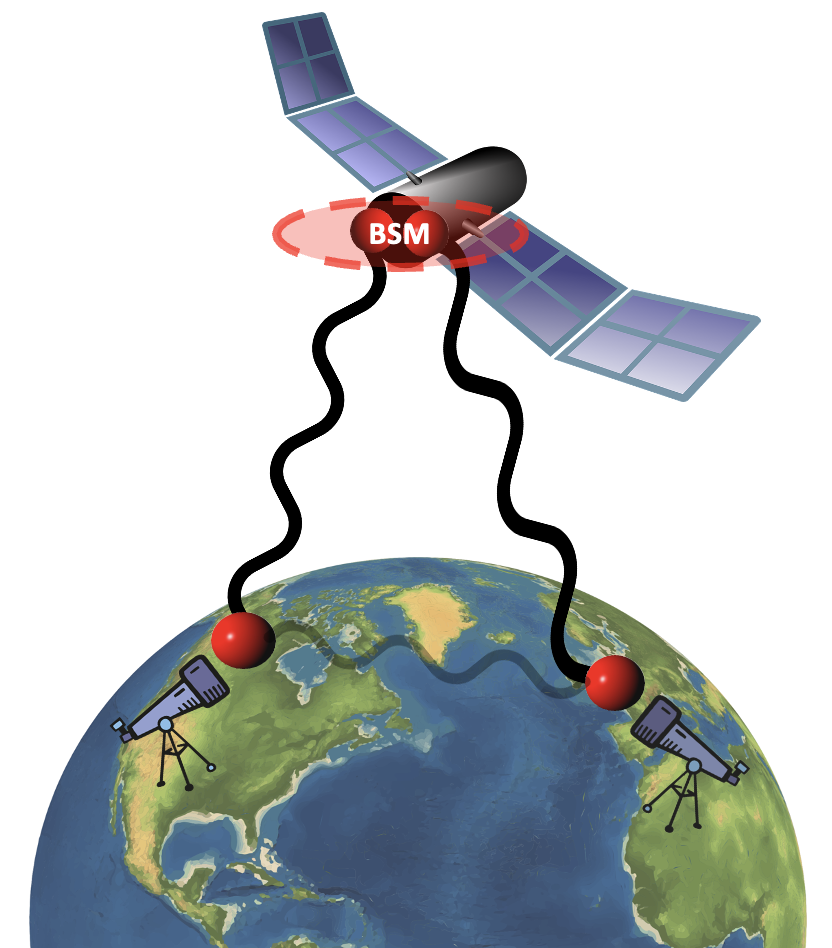}
    \caption{Two independent-downlinks can be used to share entanglement between ground-stations. In this scenario, quantum memories would allow for asynchronous preparation of the two downlinks and hence provide significant improvement in network scales with the forecasted increase in the coherence time of quantum memories. At the same time, single downlinks are less lossy (both links are created independently) and require smaller acquisition times for the same network size and quality. This puts smaller constraints on the memory coherence time as well. On the other hand, sending good quantum memories to space poses considerable scientific and engineering challenges. Furthermore, entanglement swapping is probabilistic (50\% with linear optics) with current technology. A trade-off between these aspects makes Bell violation shadows a useful metric for performance comparison.}
    \label{fig:scenario3}
\end{figure}

\begin{figure}
    \centering
    \includegraphics[width=\linewidth]{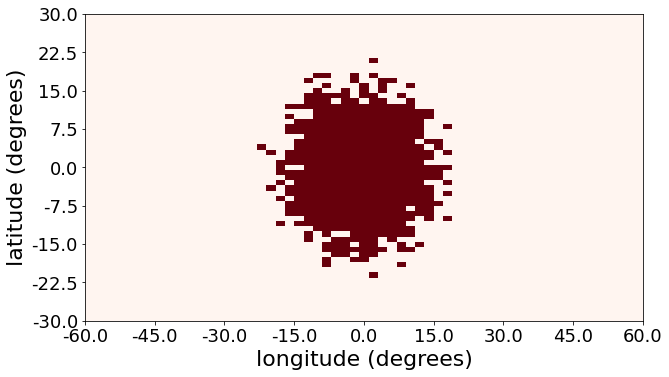}
    \includegraphics[width=\linewidth]{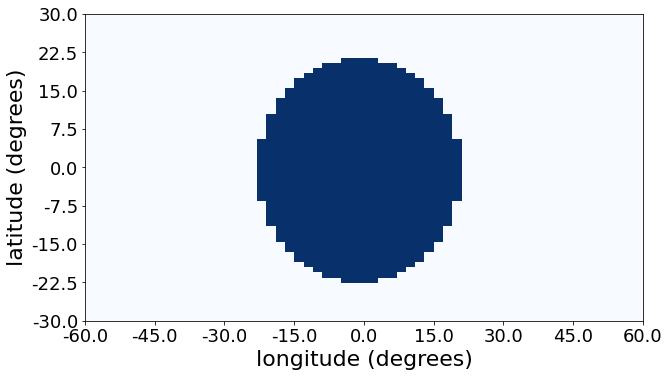}
    \includegraphics[width=\linewidth]{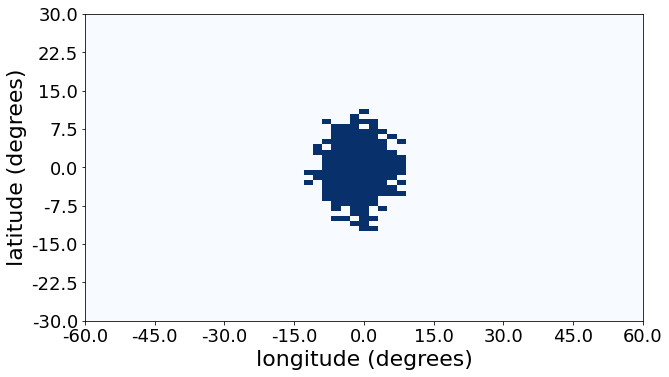}
    \caption{1-$\sigma$ Bell violation shadows relevant for quantum networks. (Top) Double-downlink scenario for quantum memories with 10 ms of coherence time (equivalent to allowing 10 ms of acquisition time in our simulations) and 1 kHz noise levels. (Middle) Two independent single-downlink scenario with entanglement swapping probability $p_{\rm sw} = 0.9$ and similar noise levels as the double downlink case gives larger and denser shadow (in fact, equal to the full visibility shadow). Thus, if swaps can be performed with high success probability the use of two independent downlinks is a better scheme for quantum networks. On the other hand, for $p_{\rm sw} = 0.75$ (bottom), the single downlink implementation gives a smaller coverage area, even though the effective loss is smaller, compared to the double-downlink case. In fact, for 1 kHz noise and 100 Hz dark count level the two independent single-downlink scenario completely fails for $p_{\rm sw} \leq 0.72$, i.e., coverage area for successful Bell violation is zero.}
    \label{fig:q_net_compare}
\end{figure}

Let us now consider two different quantum network scenarios using the Bell violation shadows, namely we compare a scenario with two independent-downlinks to the double-downlink scenario. Using a double-downlink setting, an entangled state is directly created between two ground-stations. Therefore, as mentioned above, the Bell violation shadow size is a direct measure of the scale of the quantum network in this scenario. On the other hand, for two independent-downlinks scenario, two separate downlinks must be first created from a satellite to  two ground-stations, within the coherence time of the memories on the satellite and on the ground (whichever is worse, generally the ones onboard satellites would have smaller coherence time). Then an entanglement swapping operation must be attempted between the two memories onboard the satellite, creating an entangled state between the two ground stations (see Fig.~\ref{fig:scenario3}). Using linear optics, entanglement swapping only succeeds at 50\% probability; with more complex technologies such as solid-state memories, this probability can be increased \cite{memory_review, Ruf2021} to close to 100\%. At the same time, for a given acquisition time, each single downlink Bell violation shadow is larger than the double downlink shadow, because of the lower total loss. Thus, there exists a trade-off between the two scenarios, in the latter, loss is lower but entanglement swapping---a probabilistic process---is needed, whereas in the former, loss is higher but swapping is avoided. It should be also noted that the integration of any technology to a satellite comes with its own set of challenges. Thus, it is important to assess whether using the low loss two independent single-downlink protocol would have any benefits with the added requirement of performing onboard entanglement swapping using quantum memories or linear optics. In Figure \ref{fig:q_net_compare} we illustrate such an assessment. In the top plot, we show the Bell shadow for a double-downlink scenario. In the middle plot, we show the Bell shadow obtained from two independent
single-downlinks using a high entanglement swap probability (90\%). The latter is larger and more dense than the double-downlink shadow, indicating an advantage of using quantum memories on the satellite. On the other hand, in the bottom plot we show that, as the probability of successful swaps $p_{\rm sw}$ goes down to 75\%, the single-downlink scenario performs worse (smaller and sparser shadow) than the double-downlink case. In fact, the Bell shadow for the two independent single-downlink scenario completely vanishes for $p_{sw} \leq 72\%$. For a fair comparison, we have kept the noise levels and acquisition times the same for the two scenarios.

Finally, to extend the network scale, entanglement must first be distributed and verified between several neighboring pairs of nodes within the coherence time of the quantum memories and then entanglement swapping must be performed at nodes of common connection. Such common intermediate nodes between two distant nodes are called quantum repeaters \cite{repeater_review}. Bell violation shadows can be used to decide how far from each other such repeaters must be placed (having too many repeaters can deteriorate the networks' figures of merit, since swaps are probabilistic). 

Furthermore, Bell shadows can be employed to assess the usefulness of a satellite constellation design. In a double-downlink scenario with no entanglement between satellites, ground stations must act as repeaters, whereas in a single-downlink scenario with entanglement links between satellites, the satellites themselves can act as repeaters. In the former case, to connect two far-apart cities (not falling within one shadow), the union of the Bell shadows of all the satellites must provide a continuous shadow path between them, such that swaps can be performed by ground stations falling in regions where two shadows (of different satellites) intersect. In the latter case, with entanglement links between satellites, each city merely needs to be within the shadow of one satellite and, as long as two satellites are visible to each other---either directly or transitively by means of other satellites in the constellation---ground stations within the shadows of different satellites can connect.
\begin{figure*}[ht]
    \centering
    \hspace{2cm}
    \includegraphics[width=0.3\linewidth]{Bell_shadows/dd.png}  
    \hspace{1cm}    \includegraphics[width=0.5\linewidth,height=0.37\linewidth]{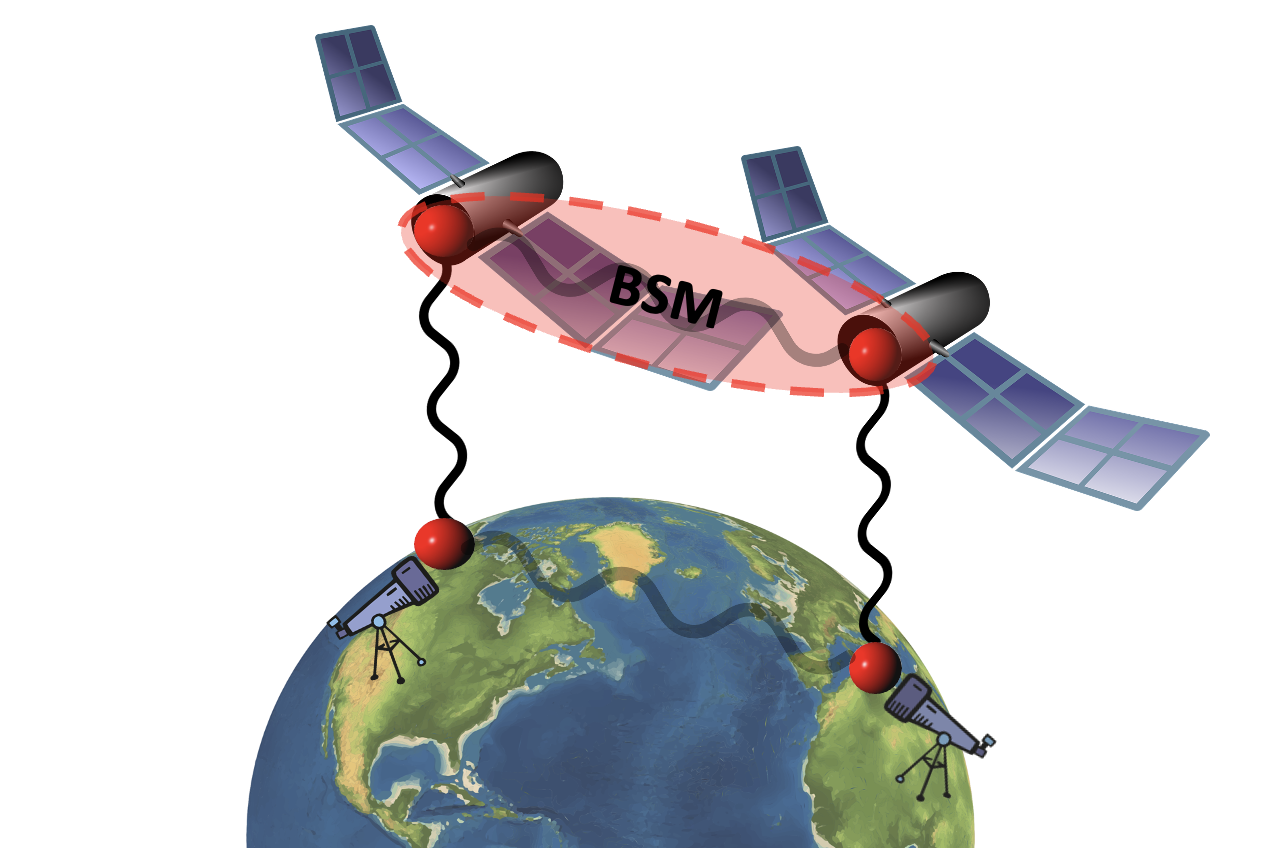}

    \includegraphics[width=0.45\linewidth]{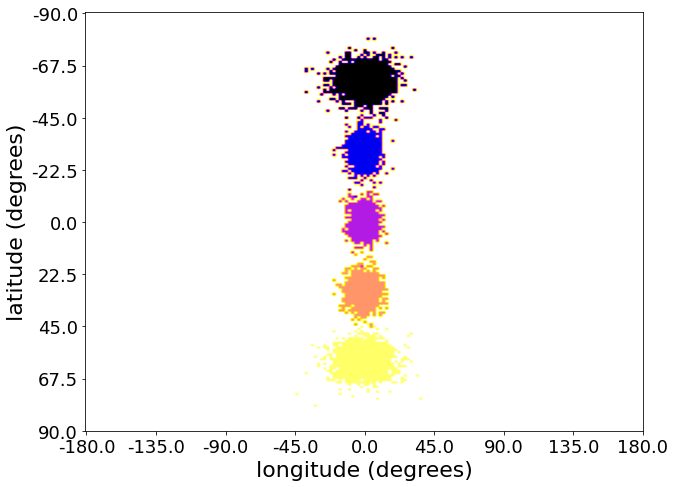}
    \includegraphics[width=0.45\linewidth]{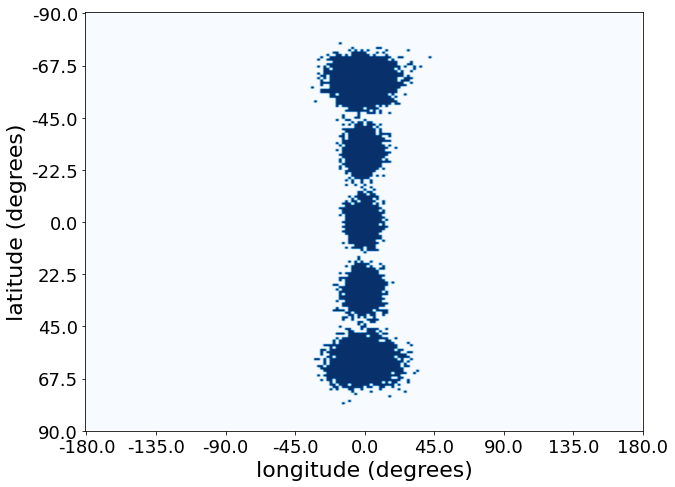}
    
    \caption{Comparison of two satellite-based quantum network schemes using Bell shadows. (Top-left) The double-downlink scenario described in section \ref{sec:Intro} and redrawn here for completeness, where an entangled photon pair is directly shared between two ground stations. (Top-right) Two independent downlinks from different satellites are combined using the entanglement swapping protocol (includes a Bell state measurement (BSM) and classical communication). For simplicity, the satellite-satellite entanglement distribution is considered to be continuously available and the swapping protocol is considered to have high success probability ($p_{\rm sw} = 0.9$).
    (Bottom left) Disconnected shadows (ground stations in different color shadows cannot share verifiable entanglement) in the double-downlink case imply a smaller network coverage with a larger coherence time requirement of 10 ms compared to (Bottom right) the swapping based scenario, which leads to a still patchy but larger Bell violation shadow for just 1 ms of coherence time for the quantum memories. In this case, the patchiness could be resolved by adding more satellites to the orbit. This  different from the double-downlink scenario, where shadow size (of single color) can only be increased by increasing the coherence time of the memories.}
    \label{fig:q_net_compare_constellation}
\end{figure*}

In Figure \ref{fig:q_net_compare_constellation} we show Bell shadows for a constellation of satellites in the same polar orbit. We use a constellation of 10 equally spaced satellites forming a polar ring at 500 km (LEO) altitude. In the double-downlink scenario, even with 10 ms coherence time for quantum memories, we get a set of disconnected shadows and hence the network coverage is roughly continental scale. Two ground stations falling in different color shadows cannot share verifiable quantum correlations. 
On the other hand, for the single-downlink case, we assume that entanglement between neighboring satellites is available deterministically.  As discussed above, if in this case satellites act as repeaters, the coherence time requirement can be considerably lowered. This leads to a patchy but much larger network. For example, if two cities under the orbit that are on opposite sides of the Earth, each fall simultaneously within a patch of the Bell shadow, we could get robust entanglement between them. To make this shadow continuous we will need to add more satellites in the orbit, and such an addition will allow global scale entanglement distribution. In both cases, we have assumed a background noise rate of 10 kHz and for the single link case an additional dark count rate of 100 Hz. We assume here that the shadows move/change negligibly within a coherence time, which is definitely true for current memories with a reasonable efficiency that have around 1-10 ms of coherence time \cite{memory_review,Ruf2021}.
Again, as shown for the previous applications, the above analysis is an illustration of the versatility of Bell violation shadows. 

\section{Conclusions and future work}
\label{sec:conclusion}
In this article, we have proposed a new metric to gauge the quality and scale of satellite-based entanglement distribution protocols: Bell violation shadows. These shadows not only incorporate the effects of loss and noise in the satellite-ground station communication channels dynamically and in real-time, but also take into account the effects of finite statistics, a limitation faced by most near-term satellite-based quantum protocols. We first show the dependence of the shadow size and density (sparseness) on the  parameters  of the protocol, such as noise and loss levels, data acquisition time, etc., in order to establish the faithfulness of our metric and to develop intuition about its behaviour in different scenarios, such as double downlinks, independent uplinks or downlinks, etc. We then go on to show the versatility of Bell violation shadows in assessing the performance of a wide class of quantum protocols, such as quantum key distribution and repeater-based quantum networks. We find that these shadows can help in the design and optimization of hardware on-board the satellite and also answer questions about constellation design. As examples of such use-cases, we compare different satellite-based entanglement distribution protocols, as shown in Figs.~\ref{fig:scenarios}, \ref{fig:scenario3} and \ref{fig:q_net_compare_constellation}, and show that different protocols become advantageous in different parameter regimes. 

The main aim of this work is to introduce a quantitative and versatile metric for the evaluation of different satellite-assisted or space-based quantum protocols. Thus, for simplicity we have ignored other complex effects like the effect of non-circular satellite orbits, atmospheric turbulence, cloud coverage, etc. 
Including these effects, along with a more detailed noise and loss model,  would be a natural extension of our current work. An extra improvement would be the inclusion of relativistic effects, which can affect the translation of the polarization measurements on the satellites and ground-stations to a common frame.  Effects of classical communication (CC) delays required for heralded entanglement generation and distillation protocols \cite{cc_policy} can also be incorporated in our simulations in the following way. The source ebit rate can be set equal to the inverse of the CC-time between the two ground stations or ground-station satellite pair performing the Bell test. Therefore, parties that are farther apart, not only have the detriment of higher loss but also of a lower source rate. Similarly, to introduce the effect of decoherence during the ebit fly-time and classical communication lag, one can assume a simple but realistic Pauli model of noise in the quantum memories. This can incorporate both the effects of dephasing and amplitude damping, these being the primary sources of errors in state-of-the-art quantum memories.
Such fully dynamic simulations (satellite, ground station motion and time evolution of quantum states) would pave the way to establish Bell violation shadows as a tool for more complex tasks such as optimization of the entanglement distribution scheduling problem in satellite-based scenarios \cite{khatri_policy, optimal_policy}. These extensions will also form part of future work.

\begin{acknowledgements}
This study was funded by the RCS program of  Louisiana Boards of Regents through the grant LEQSF(2023-25)-RD-A-04, and by the NSF grants PHY-2409402 and PHY-2110273, by funds from Xairos Inc., and by the Hearne Institute for Theoretical Physics at Louisiana State University. SH also acknowledges support from the Army Research Office Multidisciplinary University Research Initiative (ARO MURI) grant number W911NF2120214.\end{acknowledgements}

\appendix
\section{Calculation of probability distribution for the CHSH number}
\label{app:A}

In this appendix, we provide some quantitative details of the difficulty in analytically computing the probability distribution of the CHSH number in the entanglement distribution protocol under consideration in this article, and compare it with the simulation-based approach we follow.

As described in the main text, the CHSH number is defined as:
\begin{eqnarray}
    S = E(a_1,b_1) + E(a_1,b_2) - E(a_2,b_1) + E(a_2,b_2),
    \label{eqn:CHSH}
\end{eqnarray}
where $E(a_i, b_j)$ is the expectation value of the product of measurement outcomes ($\pm 1$ for the two outcomes, respectively), when measurements by Alice and Bob ($a$ and $b$) are made in the $a_i, b_j$ polarization bases, respectively ($i,j \in \{1,2\}$).

Consider that photon-polarization measurements are made for a finite acquisition time $t_{\text{acq}}$, and the CHSH number is calculated using these measurements. We call this one round or one run of the Bell test. Since ebit production from an entanglement source (e.g., SPDC) is a probabilistic process, the exact number of photons received in any run of the test is random, and are modeled as following a Poisson distribution in our simulations, with some constant rate. The possible values of $S$ for each ebit pair measurement are $\pm 1$. The probabilities associated with these outcomes, $p_1$ and $p_{-1}$, are known via the Born rule if the quantum mechanical state shared between the two parties is known. Also, $p_1 + p_{-1} = 1$. 

Say $n$ ebit pairs are detected; then, this run of the Bell test generates an $n$-dit string of outcomes like $(+1)(+1)(-1)(+1)(-1)\ldots$ (with $n$ terms), where the probability of occurrence of such an $n$-dit term is a product of the form $p_1^k p_{-1}^l$. The probability that the CHSH number after $n$ ebits are detected is equal to $S$ is then given by:
\begin{eqnarray}
    P(S,n) = \sum_{\substack{
        k,l = 1 \\ 4(k-l)/n = S,\\ k+l = n}}^{n} p_1^k p_{-1}^l,
        \label{eqn:p_sn}
\end{eqnarray}
where we have assumed that each pair of bases is chosen $n/4$ times. In other words, we must sum over all outcomes where the average value of the CHSH number comes out to be $S$, conditioned on the fact that $n$ ebits are detected.

It is possible to evaluate this sum since it is a geometric series. Thus, the probability distribution of the CHSH number in a Bell test with a finite number of measurements can be written as: 
\begin{eqnarray}
    P(S) = \sum_{n=1}^{\infty} p_n(\bar{n}) P(S,n),
    \label{eqn:p_s}
\end{eqnarray}
where $p_n(\bar{n})$ is the probability of detecting $n$ ebit pairs in a single run when the mean number of ebits detected over a large number of runs is $\bar{n}$, which is typically a Poisson distribution. $P(S)$ is the quantity of interest since the tails of this distribution give the total probability of a successful Bell test:
\begin{eqnarray}
    P(|S|>2) = \int_{-4}^{-2} P(S) \, S \, dS + \int_{2}^{4} P(S) \, S \, dS.
    \label{eqn:success_bell}
\end{eqnarray}

This probability of success depends on $\bar{n}$, the probability distribution of ebit pair detection, and the quantum mechanical state of the ebit pairs. The latter includes the information about loss and noise when the ebits are distributed via a quantum channel. 

As also shown in Ref.~\cite{Elkouss2016}, expressions like Eq.~\ref{eqn:p_sn}, a hypothesis test, can also be performed quantifying the success probability as a function of value of $S$ obtained after $n$ measurements using a P-value.

If the ebits produced by the source are perfect Bell pairs (e.g., $\ket{\psi^+}$), loss and depolarizing noise in the channel lead to a final average state, which is a Werner state:
$$
\rho(w) = \left(1-\frac{w}{4}\right) \ket{\psi^+}\bra{\psi^+} + \frac{w}{4} \mathbf{I_2},$$
where the parameter $w$ is function of loss and noise rates, and $\mathbf{I_2}$ is the two-qubit identity state.

The following difficulties appear in applying this line of thought:
\begin{enumerate}
    \item In a dynamical setting, both loss and noise vary considerably over time. It would be quite cumbersome to perform the sums in Eqs.~\ref{eqn:p_sn}, \ref{eqn:p_s}, and \ref{eqn:success_bell} efficiently for such a setting, especially when simulating a large number of satellites and ground stations. 

    In contrast, the simulation-based approach we follow in this article independently simulates the source, loss, and noise photon-by-photon, rather than incorporating them implicitly into the density matrix. This eliminates the need to perform the above-mentioned sums and provides a more realistic estimate of the CHSH number.

    \item The evaluation of $P(S,n)$ requires knowledge of the quantum mechanical state of the distributed state after it has passed through a lossy and noisy channel. Since loss and noise in the channel are themselves typically Poisson processes, the Werner state (or other forms of average density matrices) is only a valid description of the measurement statistics in the asymptotic case of infinite measurements because the Born rule itself applies to infinite copies of the state. Again, the simulation-based approach we follow is more suited to these purposes, as it automatically includes the effects of finite statistics.
\end{enumerate}

\bibliography{main}
\end{document}